\documentclass[preprint,numbers]{elsarticle}
\usepackage{graphicx} 
\usepackage{amsmath}
\usepackage{natbib}
\usepackage{bm}
\usepackage{lipsum}
\usepackage{amssymb}
\usepackage{color}
\usepackage[]{graphicx}
\usepackage{subcaption}

\newcommand{\leri}[1]{\left(#1\right)}

\title{Turbulent transport regimes in the presence of an X-point magnetic configuration}
\author[a1]{Fabio Moretti}
\author[a1]{Nakia Carlevaro}
\author[a1]{Francesco Cianfrani}
\author[a1,a2]{Giovanni Montani}
\address[a1]{ENEA, Nuclear Department, C.R. Frascati, Via E. Fermi 45, 00044 Frascati (Roma), Italy}
\address[a2]{Physics Department, ``Sapienza'' University of Rome, P.le Aldo Moro 5, 00185 Roma, Italy}

\begin{document}
\begin{abstract}
    We analyze the transport properties of the two-dimensional electrostatic turbulence characterizing the edge of a Tokamak device from the study of test particles motion (passive fluid tracers) following the $\mathbf{E}\times\mathbf{B}$ drift. We perform statistical tests on the tracer population in order to assess both the magnitude and the main features of transport. The role of other physical properties, such as viscosity and inverse energy cascade in the spectrum, is also considered. We outline that large scale eddies are responsible for greater transport coefficients, while the presence of an X-point magnetic field reduces the mean free path of the particles, however generating a larger outliers population with respect to a Gaussian profile. 
\end{abstract}
\maketitle

\section{Introduction}

The question concerning the achievement of an efficient fusion reaction on Earth, able to provide a quasi-illimited source of energy, is one of the most challenging perspective of physics and technology. 
The two main attempts to ignite a plasma, i.e. to reach a supra-critical stage of the fusion reactions, has followed two main lines of research: i) inertial fusion (IF)\cite{annurev:/content/journals/10.1146/annurev-fluid-022824-110008,ZHOU20171,Zhou_2024}, which recently \cite{Abu-Shawareb:24} got 1.5 net energy target gain exceeding the Lawson criterion \cite{Lawson:57} for self-sustaining nuclear fusion ignition; ii) magnetic confinement fusion (MCF), which is designed for the realization of a future reactor and is probably the most promising perspective for massive energy production \cite{Gibney:22}. These two approaches appear complementary, since they address the confinement issue with two completely different strategies (laser-induced implosion vs magnetic confinement), facing very different physical and technical problems. While for IF it is crucial to increase the efficiency of the energy transfer from lasers to plasma, in MCF the turbulent transport from the core to the edge of the plasma region provides energy and particle losses affecting performance, stability and safety of the reactor.

Our present study mainly concerns the heat and particle transport in the edge plasma region of a Tokamak device, that is relevant to understand and mitigate the violent particle and energy fluxes exiting the plasma that can potentially damage the first Tokamak wall \cite{Ham:20}. 

Actually, one of the main obstacles to develop an efficient fusion set-up is due to the so-called ``anomalous transport'' observed in the outflow of a magnetically confined plasma, significantly exceeding the expected Braginskii \cite{Braginskii:1965} or neoclassical \cite{Connor:73} predictions (the first corresponds to the standard collisional theory, while the second implies the inclusion of the magnetic geometry in determining the transport coefficient values due the low plasma collisionality). Since the turbulent behavior of the plasma is commonly believed to be responsible for the anomalous transport coefficients, the present discussion of the resulting statistical particle behavior in a Tokamak edge electrostatic turbulent scenario is in the direction of better understanding how to reduce or, at least, control this anomalous transport. 

It is worth stressing that turbulence is an important cross fertilization theme, emerging in many different fields, like climatology \cite{Franzke:22}, meteorology \cite{Sun:15}, cosmology \cite{Belinskii:92,Montani:95,Barrow:20}, but primarily in fluid dynamics, starting from the original treatment due to Kolmogorov \cite{Kolmogorov:41}, up to the famous works on the fractal \cite{Mandelbrot:74} and multi-fractal turbulence in \cite{Benzi:84,Frisch:85}.

In this respect, it is worth stressing how the properties of the turbulence are very sensitive to the operational conditions of a given system and, in particular, to its geometry (see the relevance of this point in aerodynamics of cars and planes \cite{Guo:24}), that in MCF is mostly the geometry dictated by the magnetic field configuration constraining particle motion. Despite this crucial peculiarity, changing from system to system of the same physical ingredient and \emph{a fortiori} from different phenomenological scenarios (neutral fluids, plasmas, spin glass, etc.), the turbulence features have some common basic properties. This is well-elucidated in the case of interest for the present analysis by the direct correspondence between the Euler equation \cite{Landau:87} for a neutral fluid and the vorticity advection in the electrostatic turbulence of a magnetically confined plasma \cite{Montani:22}, which is here faced in some details. 

Further ubiquitous phenomena are zonal flows \cite{Diamond:05,Diamond:11}, i.e. the spontaneous generation of azimuthally symmetric band-like shear flows observed in planetary atmosphere and magnetized plasmas \cite{Gurcan:15}. In MCF they can induce local turbulence quench and trigger the formation for transport barriers providing the transition from low- to high-confinement profiles (L-H transition) \cite{Zhao:17,Burrell:20}. Zonal flows are generated via an inverse energy cascade converting turbulent stress to sheared flow. Although we do not explicitly investigate here the role of zonal flows, that would require going beyond the adopted spatially local description of the edge plasma, we discuss the case with condensation in which a similar inverse energy cascade is postulated.

The electrostatic turbulence\footnote{We remark that, in an electrostatic model of turbulence, a non-zero magnetic field contribution is still retained as a background quantity. The term ``electrostatic'' indicates that fluctuations of the magnetic field are here neglected.}, characterizing the plasma in the edge region of a Tokamak device \cite{Scott_2007,Bufferand_2021,2019PhPl...26e2517S,Oliveira_2022,SCHWANDER2024106141,TAMAIN2016606}, is a fundamental 
ingredient to determine the transport properties in a realistic fusion configuration \cite{wesson2011tokamaks}. 
The first model able to capture some basic features of the non-linear drift response is the so-called Hasegawa-Wakatani scenario \citep{hase-waka83,hase-waka87,hase-mima18}  (see \cite{biskamp95,2023PhyD..45133774M,sym15091745,2024arXiv240509837C} for extensions), which identifies the source of the turbulent dynamics in the non-linear coupling between pressure perturbation and electric field fluctuations \cite{scott90,scott02}. 
As discussed in \cite{2023PhyD..45133774M,2024arXiv240509837C,montani-fluids2022,2024arXiv240701241M}, the two-dimensional turbulence, preserving the axisymmetry of the Tokamak configuration, plays an important role and, to some extent, is 
an attractor for the three-dimensional non-linear dynamics.

Here we investigate the influence that a two-dimensional turbulent plasma has on passive fluid tracers moving inside it according to the $\mathbf{E}\times \mathbf{B}$ drift velocity, 
following the same line of 
investigation started in \cite{2023JPlPh..89a9008S}. The peculiar and interesting feature of our analysis consists in the comparison of different physical settings, 
i.e. the slab magnetic configuration versus the presence of an X-point profile. 
Furthermore, we also distinguish two different turbulent regimes, corresponding to the presence in their spectral morphology of 
a condensation phenomenon 
(associated to an inverse energy cascade \cite{10.1063/1.1762301,Kraichnan_1975}) or in its 
absence (when a direct enstrophy cascade dominates). 
The slab case is also investigated in the limit of an ideal plasma, in order to understand the 
role of the thermodynamical equilibrium reached by the system in this case, as discussed in \cite{10.1063/1.861243}. 
Such an ideal limit will also offer us the possibility of a comparison of the diffusion coefficient we are able to recover from the tracer evolution and the 
corresponding theoretical value 
provided by the so-called 
$K-\epsilon$ model \cite{JONES1972301,https://doi.org/10.1002/ctpp.201610033,BASCHETTI2019200}.

Specifically, in all these cases, we develop the tracer dynamics (i.e. charged particles, which can also be thought as plasma constituents) under the influence of the $\mathbf{E}\times \mathbf{B}$ drift flow, defined by the non-linear dynamics of 
the electric field living in the plasma. 
Then, we plot the 
mean square displacement (MSD) of the particles with respect 
to their initial position versus time. 
The comparison of the MSD evolution in the different physical situations allows us to determine the different transport regimes, within a single scenario and 
in comparison to other choices of 
the field set-up. 

The manuscript is organized as follows: in Section \ref{sec2}, we present the model of electrostatic turbulence on which we base our analysis, together with its reduced version in the toroidal field (slab) case; in Section \ref{secsetup}, we describe the methods implemented to study the tracers transport; in Section \ref{sec4}, we display the results obtained. Conclusions are drawn in the final section. 

\section{Model of electrostatic turbulence with an X-point magnetic configuration}\label{sec2}

Here we provide a presentation of the model describing the turbulent dynamics due to the non-linear low frequency drift response, in the presence of a background magnetic configuration able to mimic the features of the X-point region of a Tokamak device. We start by illustrating the local magnetic equilibrium on which we will subsequently set the analysis of electrostatic fluctuations. Neglecting toroidal curvature effects, we adopt a set of Cartesian coordinates $(x,y,z)$ centered in the X-point, i.e. the origin of our flat grid corresponds to the null of the poloidal magnetic field. The coordinates $(x,y)$ span the poloidal plane, while $z$ is taken along the toroidal direction. The magnetic field $\mathbf{B}$ is expressed as
\begin{equation} \label{bgmagnfield}
    \mathbf{B}= B_p y \mathbf{\hat{e}_x}+B_p x \mathbf{\hat{e}_y}+B_t\mathbf{\hat{e}_z},
\end{equation}
where $B_p$ and $B_t$ are two constants accounting for the strength of the poloidal and toroidal magnetic field components, respectively. By introducing the poloidal length scale $L_p$, the request of having the magnetic field toroidal component as the dominant contribution is translated in the condition $B_t \gg B_p L_p$, which has to be satisfied in all points in which the magnetic configuration \eqref{bgmagnfield} results valid. Further, it is possible to define the directional versor $\!\!\hat{\,\,\textbf{b}}=\frac{\mathbf{B}}{B}$, in which $B$ represents the coordinate-dependent magnitude of the magnetic field, namely
\begin{equation}
    B(x,y)=B_t \sqrt{1+\frac{B_p^2\leri{x^2+y^2}}{B_t^2}}.
\end{equation}
We stress that we always speak of parallel and perpendicular directions referring to the versor $\!\!\hat{\,\,\textbf{b}}$. For instance, by setting $B_p=0$ it turns out that the parallel direction corresponds to the toroidal one (namely the $z$ axis), whereas by taking $B_p\neq 0$ the versor $\!\!\hat{\,\,\textbf{b}}$ acquires a non-null component on the poloidal plane. Moreover, the definition of the directional versor $\!\!\hat{\,\,\textbf{b}}$ allows for an unambiguous splitting of the gradient operator $\boldsymbol{\nabla}$ in a parallel and a perpendicular contribution, i.e. $\boldsymbol{\nabla}=\boldsymbol{\nabla}_{\perp}+\boldsymbol{\nabla}_{\parallel}$, where the parallel operator is defined as $\boldsymbol{\nabla}_{\parallel}=\!\!\hat{\,\,\textbf{b}}(\!\!\hat{\,\,\textbf{b}}\cdot\boldsymbol{\nabla})$ while the perpendicular component is simply recovered from $\boldsymbol{\nabla}_{\perp}=\boldsymbol{\nabla}-\boldsymbol{\nabla}_\parallel$.

Now we enumerate the hypotheses on which our model of electrostatic turbulence is based. We deal with a Hydrogen-like plasma, which satisfies the conditions of homogeneity and isotropy. We adopt a two-fluid description, assuming quasi-neutrality (having observed that the turbulence length scale is always much greater than the Debye length of the plasma we consider), i.e. we set $\mathcal{N}_i=\mathcal{N}_e= \mathcal{N}$, being $\mathcal{N}_{i/e}$  the ion/electron number density. This hypothesis, together with the further assumption of equal temperature $T$ for ions and electrons, immediately implies that we can consider a unique pressure for the two fluids, i.e. $p_i=p_e=p=\mathcal{N} K_B T$, where we use the standard symbol $K_B$ to denote the Boltzmann constant. Moreover, we take the ion polarization drift as the dominant contribution to the orthogonal motion and we neglect both diamagnetic effects and parallel ion velocity. Those just listed are the ordinary assumptions that are necessary to derive the 3D Hasegawa-Wakatani model \cite{biskamp95} describing electrostatic edge turbulence. Our original contribution is condensed in the last hypothesis, namely we neglect the gradients of the background quantities. Hence, in this work, we consider $\mathcal{N}=const.$, $p=const.$ and $T=const.$, so that all the dynamical variables are perturbations denoted by a symbol $\delta$.

Let us now turn to the construction of the model. The first dynamical equation we consider is the unique (due to the quasi-neutrality assumption) continuity equation for both ions and electrons
\begin{equation}\label{continuity}
	\frac{d \delta \mathcal{N}}{dt} 
    -\mathcal{D}\nabla^2_{\perp} \delta \mathcal{N} = 
	\frac{1}{e}\boldsymbol{\nabla}_{\parallel}\cdot \delta\textbf{J}_{\parallel}
	\,,
\end{equation}
where a diffusive term, controlled by the coefficient $\mathcal{D}$, has been introduced to take into account different regimes of number density transport. The symbol $\delta\textbf{J}_{\parallel}$ represents the fluctuating parallel current density and $e$ is the electron charge (we adopt Gaussian units). In this work we consider the $\textbf{E}\times \textbf{B}$ drift velocity as the dominant contribution affecting the advective derivative, i.e. $d/dt =  \partial_t+ \delta\textbf{v}_{\perp}^{e}\cdot\boldsymbol{\nabla}$, with
\begin{align}
	&\delta\textbf{v}_{\perp}^{e} = \frac{c}{B^2} \textbf{E}\times\textbf{B} = \frac{c}{B}\;\!\!\hat{\,\,\textbf{b}}\times \boldsymbol{\nabla}_\perp\delta\phi\,=\nonumber\\
 &=\frac{c}{B^2} \big [ 
 \leri{B_p x\partial_z\delta\phi-B_t\partial_y\delta\phi}\hat{\textbf{e}}_x+
 \leri{B_t\partial_x\delta\phi-B_p y\partial_z\delta\phi}\hat{\textbf{e}}_y+ \label{advection} \\
 &\qquad \qquad +\leri{B_p y\partial_y\delta\phi-B_p x\partial_x\delta\phi}\hat{\textbf{e}}_z \big ]\,\nonumber,
\end{align}
being $c$ the speed of light and $\delta\phi$ the electric field potential. Next, we write down the momentum balance equation for ions along the perpendicular direction
\begin{equation}
\frac{d\,\delta\textbf{v}_{\perp}^{i}}{dt} = 
	 \frac{e}{m_i}\left( -\boldsymbol{\nabla}_{\perp}\delta\phi + 
	\delta\textbf{v}_{\perp}^{i}\times \textbf{B}/c\right) 
	+ \nu \nabla^2_{\perp}\,\delta\textbf{v}_{\perp}^{i}
	\, , 
 \label{ionmombalance}
\end{equation}
where $m_i$ and $\nu$ are the ions mass and specific viscosity (here we neglect the parallel component of the viscous stress). The latter can be explicitly calculated as
\begin{align}
    \nu=\frac{1}{m_i\mathcal{N}}\frac{(3/10)\;\mathcal{N} K_B T}{\Omega_i^2 \tau_{ii}}\,,
    \qquad\qquad
    \tau_{ii}=\frac{3\sqrt{\mathcal{N}}(K_B T)^{3/2}}{4\sqrt{\pi}\;e^4 \mathcal{N} \mathrm{ln}\Lambda_{ii}}\,,
\end{align}
in which $\Omega_i=eB_t/c m_i$ is the ion gyro-frequency due to the toroidal magnetic field, the Coulomb logarithm has been set to $\mathrm{ln}\Lambda_{ii}=21$ and we have exploited the quasi-neutrality condition. Due to the fact that we expect $\Omega_i$ to be much greater  of the turbulence time-scale inverse, we can assume $\delta\textbf{v}_{\perp}^{i} = \delta\textbf{v}_{\perp}^{e} + \tilde{\textbf{v}}_{\perp}^{i}$, with $\tilde{\textbf{v}}_{\perp}^{i}$ a small correction. The latter can be calculated from a first-order expansion of \eqref{ionmombalance}, resulting in
\begin{equation}
	\tilde{\textbf{v}}_{\perp}^{i}=\frac{cm_i}{Be}\Big( 
	\frac{d\,}{dt} - \nu\nabla^2_{\perp}\Big)(\,\!\!\hat{\,\,\textbf{b}}\times \delta\textbf{v}_\perp^e) =-\frac{c^2m_i}{Be}\Big( 
	\frac{d\,}{dt} - \nu\nabla^2_{\perp}\Big)\frac{1}{B} \boldsymbol{\nabla}_\perp\delta\phi
	\,.
\end{equation}
Hence, by splitting the charge conservation equation, namely $\boldsymbol{\nabla}\cdot\delta\textbf{J}=0$, in its parallel and perpendicular contributions, we obtain
\begin{equation}\label{chargecons}
\boldsymbol{\nabla}_\perp\cdot\left(
\frac{1}{B}\frac{d\,}{dt}\frac{1}{B} \boldsymbol{\nabla}_\perp \delta\phi\right)- \nu
\boldsymbol{\nabla}_\perp\cdot\left(\frac{1}{B}\nabla_\perp^2\frac{1}{B} \boldsymbol{\nabla}_\perp \delta\phi\right)=
\frac{1}{\mathcal{N}c^2m_i}\boldsymbol{\nabla}_{\parallel}\cdot\delta\textbf{J}_{\parallel}\,, 
\end{equation}
having noticed that the orthogonal term can be written as 
$\delta\textbf{J}_{\perp} \equiv \mathcal{N} e(\delta\textbf{v}_{\perp}^{i} -\delta\textbf{v}_{\perp}^{e})=\mathcal{N} e \,\tilde{\textbf{v}}_{\perp}^{i}$. The electron momentum balance along the parallel direction results in the generalized Ohm law
\begin{align} \label{genohm}
\delta\textbf{J}_{\parallel} = \frac{\sigma}{\mathcal{N}e} \boldsymbol{\nabla}_{\parallel}\delta p -\sigma \boldsymbol{\nabla}_{\parallel}\delta\phi\,,
\end{align}
where we introduced the parallel conductivity $\sigma\equiv1.96\, \mathcal{N} e^2/m_e \nu_{ei}$, with $\nu_{ei}$ the electron-ion collision frequency and $m_e$ the electron mass. We stress that, given the assumptions adopted in this analysis, the fluctuating pressure $\delta p$ can be expressed in terms of the number density via the equation $\delta p = K_B T\delta \mathcal{N}$.

The set \eqref{continuity}, \eqref{chargecons} and \eqref{genohm} constitutes the system of dynamical equations characterizing our model. We introduce adimensional coordinates $\tau\equiv \Omega_i t$, $u\equiv(2\pi/L_p) x$, $v\equiv(2\pi/L_p) y$, $w\equiv (2\pi/L_t) z$ ($L_t$ being the toroidal spatial scales, with $L_t\gg L_p$), together with the parameters 
\begin{align}
\varepsilon=\frac{B_p}{B_t}\,\leri{\frac{L_p}{2\pi}}, \qquad \gamma(u,v)=\sqrt{1+\varepsilon^2(u^2+v^2)}    
\end{align}
accounting for the small ratio of the poloidal component with respect to the toroidal one and the X-point geometry, respectively. The dimensionless version of the parallel gradient vector is obtained from $\textbf{D}_{\parallel}=(L_p/2\pi)\boldsymbol{\nabla}_{\parallel}$ and its explicit expression reads
\begin{align}
\textbf{D}_{\parallel}=\Big( 
\frac{\varepsilon}{\gamma}(v\partial_u+u\partial_v)+ \frac{L_p/L_t}{\gamma}\,\partial_w
\Big)\!\!\hat{\,\,\textbf{b}}\,.
\end{align}
From the latter it is immediate to obtain the perpendicular gradient vector $\textbf{D}_{\perp}$ and the Laplacian operators $D_{\parallel}^2$ and $D^2_{\perp}$, whose lengthy and cumbersome explicit expressions are not reported here to avoid unnecessary complexity. Setting $\Phi\equiv e\,\delta\phi /K_BT$, $\bar{\mathcal{N}}=\delta\mathcal{N}/\mathcal{N}$ and $\textbf{Y}_{\parallel} =(2\pi/L_p)\delta\textbf{J}_{\parallel}/ \mathcal{N}e\Omega_i$, we summarize the dynamical equations of our model in dimensionless form as follows
\begin{align}
\frac{d}{d\tau}\bar{\mathcal{N}}-\bar{\mathcal{D}} D^2_{\perp} \bar{\mathcal{N}} &= 
\boldsymbol{D}_{\parallel}\cdot\textbf{Y}_{\parallel}\;,\label{eqn}\\
\alpha_1\boldsymbol{D}_\perp\cdot\left(
\frac{1}{\gamma}\frac{d\,}{d\tau}\frac{1}{\gamma}\boldsymbol{D}_\perp \Phi\right)
-\alpha_1\alpha_2
\boldsymbol{D}_\perp\cdot\left(\frac{1}{\gamma}D_\perp^2\frac{1}{\gamma}\boldsymbol{D}_\perp \Phi\right)&= 
\boldsymbol{D}_{\parallel}\cdot\textbf{Y}_{\parallel}\;,\label{eqvort} \\
 \alpha_3 (D_{\parallel}^2\bar{\mathcal{N}}  -D_{\parallel}^2\Phi)&=\boldsymbol{D}_{\parallel}\cdot\textbf{Y}_{\parallel}\,.\label{eqohm}
\end{align}
The explicit expressions of the dimensionless constants appearing in our equations are here reported:
\begin{align}
\alpha_1= \rho_i^2\leri{\frac{2\pi}{L_p}}^2,\quad
\alpha_2=\frac{\nu}{\Omega_i}\leri{\frac{2\pi}{L_p}}^2, \quad
\alpha_3=\frac{v_A^2 \rho_i^2}{\eta_B \Omega_i}\leri{\frac{2\pi}{L_p}}^2,\quad
\bar{\mathcal{D}}=\frac{\mathcal{D}}{\Omega_i}\leri{\frac{2\pi}{L_p}}^2,
\end{align}
where $\rho_i^2=K_B T/m_i \Omega_i^2$ is the ion Larmor radius, $v_A=B_t/\sqrt{4\pi \mathcal{N} m_i}$ the Alfv\'en velocity constructed with the toroidal magnetic field and $\eta_B=c^2/4\pi\sigma$  the magnetic diffusivity.

\subsection{Reduced model: pure toroidal magnetic field}

In this section we provide a description of a reduced version of our model, reproducing the case of a constant toroidal magnetic field, thought as a predictive paradigm sufficiently close to the X-point. Indeed, by setting $B_p=0$, which in turns implies $\epsilon=0$ and $\gamma=1$, we obtain the following simplified expressions for the differential operators entering the dynamical equations of the model:

\begin{equation}
\boldsymbol{D}_\parallel \to \frac{L_p}{L_t}\partial_w \mathbf{\hat{e}_w}\,,\quad
\boldsymbol{D}_\perp \to \partial_u\mathbf{\hat{e}_u}+\partial_v\mathbf{\hat{e}_v}\,.
\end{equation}
For what concerns the temporal evolution of the fields involved in our analysis, the Lagrangian advective derivative, whose general expression can be deduced from \eqref{advection}, reads

\begin{align}
\frac{d}{d\tau}=\partial_{\tau}+\alpha_1(\partial_{u}\Phi \partial_{v}-\partial_{v}\Phi\partial_{u})\,.
\end{align}
In this simplified scheme equations \eqref{eqn} and \eqref{eqvort} result identical, provided that the diffusion coefficient is set to $\mathcal{D}=\nu$. Therefore, the model is condensed in a single equation for the vorticity, namely
\begin{align}
\partial_{\tau}D^2_{\perp}\Phi +\alpha_1\big( \partial_{u} \Phi \partial_{v} D_{\perp}^2\Phi - \partial_{v}\Phi\partial_{u}D_{\perp}^2\Phi\big)=\alpha_2\; D_{\perp}^4\Phi + \alpha_3\; D_{\parallel}^2D_{\perp}^2\Phi-\frac{\alpha_3}{\alpha_1}\;D^2_{\parallel}\Phi \,  
\label{eqvortunica}
\end{align}
and the number density is recovered from the constitutive relation
\begin{align}\label{relcost}
\bar{\mathcal{N}}=\alpha_1 D_\perp^2\Phi\;. 
\end{align}
As we showed in a previous work \cite{sym15091745}, by enforcing periodic boundary conditions and writing the electric potential in Fourier series

\begin{align}
\Phi (\tau,u,v,w) = \sum_{n,\ell,m}\varphi_{n,\ell,m}(\tau)\;e^{i(nw+\ell u+mv)}\,,\label{ccm8}
\end{align}
where $(n,\ell,m)$ are integers numbers, the dynamics described by equation \eqref{eqvortunica} naturally relaxes to a 2D configuration: every non-constant contribution along the $w$ direction, i.e. every mode with $n \neq 0$, is rapidly suppressed due to a combined effect of inverse energy cascade and dissipation. Indeed, by initializing to zero the $n=0$ mode and to random noise all the others $n \neq 0$ modes, it can be observed that, after a short ($\tau \sim 200$) time of simulation, part of the energy stored in the $n \neq 0$ modes has been transferred to the $n=0$ mode, while the rest has been dissipated by the viscous term. Hence, in this work we study the axisymmetric turbulence in the poloidal plane, by considering a solution $\Phi=\Phi(\tau,u,v)$ of the 2D restriction of \eqref{eqvortunica}, namely
\begin{equation}
\partial_{\tau}D^2_{\perp}\Phi +\alpha_1\big( \partial_{u} \Phi \partial_{v} D_{\perp}^2\Phi - \partial_{v}\Phi\partial_{u}D_{\perp}^2\Phi\big)=\alpha_2\; D_{\perp}^4\Phi.\label{tordued}
\end{equation}
It has to be remarked that this equation is equivalent to a 2D Euler equation for an incompressible fluid with a viscous term in the vorticity representation \cite{montani-fluids2022,2023PhyD..45133774M}. The equivalence becomes evident once the mapping 
\begin{equation}\label{mapping}
    \leri{v_x,v_y}\to \leri{-\partial_v \Phi,\partial_u \Phi}
\end{equation}
is implemented (here $v_x$ and $v_y$ are the Cartesian components, as referred to a coordinates system ($x,y$), of the incompressible flow velocity). Hence, in this analogy with the fluid theory, the electric potential $\Phi$ plays the role of a stream function, i.e. the curves on which $\Phi$ is constant correspond to streamlines. Moreover, having recognized this formal resemblance with the 2D Euler equation allows us to infer a number of properties of \eqref{tordued} without further analyses. First, the dynamics associated to the 2D Euler equation and, by extension, to \eqref{tordued}, results with great generality in a turbulent behavior. In addition to this, it can be demonstrated that, in the ideal inviscid case, there is an infinite number of conserved quantities associated to such equation. Indeed, when periodic boundary conditions are imposed on a 2D region of size $A$, it turns out that the specific kinetic energy
\begin{equation}
    K=\frac{1}{2A}\int d^2x \, v^2(x,y)
\end{equation}
and every other Casimir invariant constructed with a generic function of the vorticity norm $\omega=|\mathbf{\nabla}\times \mathbf{v}|$, namely
\begin{equation}
    G_g=\frac{1}{2A}\int d^2x \, g(\omega),
\end{equation}
result preserved along the dynamics. Let us focus on a specific realization of a Casimir invariant, namely the enstrophy
    \begin{equation}
    \Omega=\frac{1}{2A}\int d^2x \, \omega^2(x,y).
\end{equation}
Now, if we imagine to expand both the velocity and the vorticity fields in Fourier series, it is sufficient to invoke Parseval theorem to affirm that the conserved quantities can be calculated from
\begin{equation}
    K=\frac{1}{2}\sum_k |v_k|^2, \qquad \Omega=\frac{1}{2}\sum_k |\omega_k|^2,
\end{equation}
in which $v_k$ and $\omega_k$ are the Fourier coefficients of $v$ and $\omega$ respectively. 
In the equation above the sums are intended to be performed on every allowed value of the wavenumber $k$ but, in practice, on always handle truncated sums in which $k$ ranges between a minimum and a maximum, i.e. $k_{min}^2 \leq k^2=k_x^2+k_y^2 \leq k_{max}^2$. For instance, if we assume that the 2D region on which we are integrating Euler equation is a square with side $L$, the minimum wavenumber $k_{min}$ corresponds to $\frac{2\pi}{L}$, whereas no uncontroversial upper bound can be imposed on physical grounds in the fluid case, and the cut-off $k_{max}$ is very often connected to a mere realistic computational time limit. Nonetheless, it can be shown that $K$ and $\Omega$ are still conserved even when a description through a truncated Fourier series is performed, while many other Casimir invariants result no longer constants. In our case, we implement an expansion of the electric potential in a 2D Fourier series 

\begin{equation}\label{fourierseries}
    \Phi (\tau,u,v) = \sum_{\ell,m}\varphi_{\ell,m}(\tau)\;e^{i(\ell u+mv)}
\end{equation}
with the Fourier coefficients\footnote{In the following, the explicit time dependence of the Fourier amplitudes $\varphi_{\ell,m}$ is dropped, for the sake of simplicity.} satisfying the reality condition $\varphi_{-\ell,-m}=\leri{\varphi_{\ell,m}}^*$, where the $*$ indicates complex conjugation. The mapping \eqref{mapping} clearly shows that, in this plasma analogue of Euler equation, the (dimensionless) kinetic energy $\bar{K}$ and enstrophy $\bar{\Omega}$ are calculated via 
    \begin{equation}\label{costantimoto}
    \bar{K}=\sum_{\ell,m} \leri{\ell^2+m^2}|\varphi_{\ell,m}|^2, \qquad \bar{\Omega}=\sum_{\ell,m} \leri{\ell^2+m^2}^2|\varphi_{\ell,m}|^2.
\end{equation}
Another aspect that has to be emphasized is that a physical turbulent system governed by the 2D Euler equation can undergo a process of condensation, namely a transfer of energy towards the lowest wavenumber allowed by the geometry and the boundary conditions, corresponding to an inverse cascade of energy towards the largest spatial scale. It can be shown that the fundamental parameter separating the regimes of condensation and non-condensation is a characteristic wavenumber obtained as $k^2_c=\frac{\bar{\Omega}}{\bar{K}}$. The inverse of $k_c$ corresponds to the spatial size of the vortices that will dominate the dynamics once the transient initial phase is ended and equilibrium is reached. Therefore, a small value of $k_c$ indicates that the system will condensate, whereas a large one will lead to an equilibrium characterized by vortices of many different sizes exchanging energy without a preferred direction in the $k$ space (we remark that this occurs solely in the ideal inviscid case, given that the usual viscous term has a form $\propto k^4$ in the Fourier space, causing high-wavenumber modes to be heavily suppressed in the late stages of the system evolution). The precise value of the threshold between the two regimes can be empirically determined for each specific physical system under consideration, the former being dependent on the geometry taken into account and the choice of length and time scales used to define dimensionless coordinates. In our case, for instance, it turns out (see also \cite{10.1063/1.861243}) that condensation is achieved by choosing the initial non-zero Fourier mode such that it results $k_c^2\simeq 10$, whereas the non-condensation scenario is reproduced for $k_c^2\simeq 60$. These values for the characteristic wavenumbers are obtained by initializing the simulation with all modes to zero except for $k^2=\ell^2+m^2=5$ or $13$ in the first case and  $k^2=52$ or $85$ in the absence of condensation. The amplitudes of the initial modes are set in order to have the total energy of the system $\bar{K}=1.4 \times 10^{-3}$.

It is worth stressing that the analysis of the electrostatic 2D turbulence
in the
slab magnetic case, is in all parts isomorphic to the neutral fluid turbulence,
both in the inviscid and viscous
regimes.
This can be easily realized by observing that the $\mathbf{E} \times \mathbf{B}$ velocity is, in
such a magnetic configuration, a divergenceless flux field.
Hence, it is immediate to interpret
our turbulence (see \cite{10.1063/1.861243}) as corresponding to an Euler equation,
in which the stream function is mapped on to minus the
electric field \cite{montani-fluids2022}.
From a more phenomenological point of view, we have that the ion fluid
(responsible for the main plasma inertia) is well-described by an
incompressible (inviscid or viscous) 2D flow.
However, this relevant correspondence
between the plasma and an incompressible fluid is formally lost
when the X-point configuration is introduced. In fact, the magnetic field
geometry prevents, in such a case,
a direct comparison of the vorticity
advection equation as a real Euler equation, due to the presence of
additional contributions. Nonetheless,
near enough to the X-point, the deviation from the Euler equation becomes
small and the main
features of the plasma towards a correspondence with neutral fluids is still recovered in
the turbulence features.

\subsection{On the reliability of the reduced model}

Here, we want to briefly discuss the 
capability of the reduced model to 
capture significant features of 
the Tokamak edge turbulence, despite some apparent simplifications at the base of its definition.

First of all, we observe that the 
assumption to deal with an uniform plasma background temperature and number density is a viable picture, especially when 
their values are thought as mean values over a local real equilibrium configuration. Actually, the dynamics of the temperature is, in general, considered as a basic feature of the electrostatic turbulence, as 
confirmed by valuable versions of 
dedicated codes, see for instance 
the original version of TOKAM3X 
\cite{TAMAIN2016606}.
Actually, in the original Hasegawa-Wakatani model \cite{hase-waka83,hase-waka87}, the background density is taken exponentially varying along one of the two poloidal coordinates, so that the spatial gradient in that direction is proportional to 
the density itself. It is 
important to stress that the resulting electrostatic turbulent model is, 
however, at constant coefficients, 
i.e., given its form, the introduced spatial dependence (without a specific reference to a background equilibrium) has not a direct dynamical impact, 
except for the emergence of the 
background gradient in the drift coupling\footnote{This feature would not hold in the presence of magnetic fluctuations, since the density and pressure spatial dependence would 
enter the dynamics.}. 
Moreover, we have to stress that this 
background gradient is assumed to be negligible in the construction of the reduced model and, therefore, having 
constant values of the background 
number density and pressure 
(to be thought as local average values, 
taken near the X-point) is not a significant physical restriction. 
Regarding the negligibility of the 
background pressure gradient, we 
observe that ignoring them corresponds to eliminate the free energy source triggering the linear drift
instability. However, it is important to outline that, as discussed in
\cite{scott02,Scott_2007}, the real electrostatic turbulence regime
is not directly related to the liner triggering, while it is
a well-known self-sustained process.
By other words, the intrinsic nature of the non-linear drift response is
characterized by the self-coupling of the electric field, namely the
advection of the vorticity, together with the
non-linear coupling between the electric field and the perturbed pressure
term, i.e. the advection of the pressure fluctuations.
After the system is initialized,
the associated energy injection is responsible for the fully developed
self-sustained turbulence, independently from the original nature of
such energy content (actually, it can be originated by the saturation of a
linear drift instability, by particle or heat fluxes from the core to the
edge of
the plasma, etc.).
Since we are interested in studying the
influence of the electrostatic turbulence on particle
transport, the addressed reduced model (see \cite{2023PhyD..45133774M,sym15091745}) is
adequate to capture all the real
features characterizing the average
properties of such tracers when they
experience the field fluctuation spectrum (independently from the
details of its establishment).

Let us now turn to the discussion of one of the main features characterizing the reduced model: the constitutive relation \eqref{relcost}. We remark, firstly, that such equation, connecting 
the electric vorticity  and the number density, is not the most 
general picture. In fact, from a direct inspection  
of the dynamical equations \eqref{eqn}, \eqref{eqvort} and \eqref{eqohm}, we see that the difference of these two 
quantities, namely the potential vorticity \cite{10.1063/5.0041072,10.1063/5.0189855,2024arXiv240309911G} defined as $\Pi=\bar{\mathcal{N}}-D^2_\perp \Phi$, would verify an advective 
heat equation (we outline that this latter equation, being homogeneous, admits the null function as 
a viable solution). 
However, two main reasons justify our choice 
to consider this specific case:
i) In the reduced model, we aim at 
a certain degree of simplification of the dynamics, in which the evolution of 
one degree of freedom (here the number density) is determined by the 
dynamics of the remaining one 
(here the electric field potential). 
More general choices would not correspond to a reduction of the dynamics, 
but just to its rewriting via combined degrees of freedom.
ii) It is well-known that the 
proposed scenario offers the right 
scheme for the discussion of the 
electrostatic turbulence, simply 
expressing the number density via the 
corresponding advection of the electric field. The physical content of the 
model remains predictive since it is 
just the self-interaction of the 
electric field (say its advection) 
the fundamental non-linear ingredient of the self-sustained 2D turbulence, as clarified by the analogy with the Euler equation for 
an incompressible fluid (see the previous section). 

Finally, we comment on the assumption of dealing with an electrostatic turbulence only, when Tokamak edge physics is addressed. 
This assumption has been the starting point of all the realistic codes 
built over the last ten years, simply because this effect is always present 
and it is expected to dominate the 
dynamics of the magnetic fluctuations. This consideration finds its justification in the relative low value of 
the plasma $\beta$-parameter 
(i.e. the average ratio between the thermodynamical and magnetic pressure) in the operational settings of current 
Tokamak devices, as well as 
in many scenarios of incoming 
experiments. 
In fact, as far as the $\beta$-parameter remains around few percents, 
the magnetic fluctuations are, 
to some extent, frozen out and, 
in any case, subdominant with respect to 
the electrostatic contribution to 
the turbulent transport \cite{scott02,TAMAIN2016606}. In addition, our study of the 
tracer dynamics essentially concerns 
the $\mathbf{E}\times\mathbf{B}$ flow and the corresponding 
velocity would be affected by the presence of magnetic fluctuations 
at higher order only 
(i.e. at the order of the strength associated to the product of electric and magnetic fluctuations). 

Summarizing, the above discussion states that, despite some specific assumptions, the reduced model we investigate here is naturally able to capture 
relevant physical features of the 
Tokamak edge turbulent transport, 
especially when we are interested 
to the turbulence impact on the 
$\mathbf{E}\times\mathbf{B}$ flux of plasma constituents.

\section{Setup of the numerical analysis}\label{secsetup}

In this section we provide a description of the numerical methods implemented to analyze the transport of particles. First, we briefly report the details of the numerical integration for the electric potential, performed either on equation \eqref{tordued}, when the reduced model is considered, or on the complete set \eqref{eqn}, \eqref{eqvort}, \eqref{eqohm}, in the case of an X-point configuration. In both scenarios our domain of integration is a square in the poloidal plane of side $2\pi$, on which we impose periodic boundary conditions. The time integration is performed via a fourth-order Runge-Kutta method, with time step $h=10^{-3}$ and a total simulation time corresponding to $\tau=10^4$, the latter being much greater than the time for the system to achieve a stationary state, i.e. a quasi-steady morphology of the spectrum (of course, we refer here to the inviscid scenario). The non-linear terms characterizing the equations of our model are treated by resorting to a pseudo-spectral approach. We consider a finite number of Fourier modes $(\ell,m)$ according to the following prescriptions: the lower bound for the poloidal wavenumber $k$ is provided by the box size, i.e. $k_{min}=\frac{2\pi}{L_p}$, whereas the fluid description of plasma adopted in this work implies that the smallest spatial scale accessible must be greater than the ion Larmor radius, so that it results $k_{max}=\frac{2\pi}{\rho_i}$. Hence, we safely consider $\ell$ and $m$ ranging between $-13$ and $13$ with any possible combination taken into account, excluding the case in which both indices are null, given that the corresponding term describes a mere constant contribution. This entails a total number of independent modes equal to $364$. For what concerns the case with the X-point configuration, we adopted the same numerical scheme presented in \cite{2024arXiv240509837C} for constant and radially sheared poloidal magnetic field in addition to the toroidal one, then extended to the X-point case in \cite{cianfmont}. This approach is based on solving the equations for vorticity and perturbed number density in 3D by deriving at each iteration step the electrostatic potential using a pre-computed inverse perpendicular Laplacian. An efficient implementation of this scheme is realized in Python using the functionalities of NumPy \cite{2020Natur.585..357H} and SciPy \cite{scipy} packages. In this work, this method is restricted to 2D and to the equations of the reduced model by imposing the constitutive relation \eqref{relcost} on initial data. Typical Tokamak parameters are assumed for both the slab and the X-point cases: $T=100\,$eV, $B_t=3\,$T and $\mathcal{N}=5\times10^{19}\,$m$^{-3}$ \cite{dtt19}, hence $\Omega_i\simeq1.4\times 10^8\,$s$^{-1}$, $\rho_i\simeq0.048\,$cm and $\nu\simeq 24.9\,$cm$^2$s$^{-1}$. The poloidal length scale is set to $L_p=1$cm.

Let us now focus on the main goal of this work, namely the study of turbulent transport from the motion of passive fluid tracers. We initialize a total number $N=10^4$ of tracers randomly distributed on the square of side length $2\pi$ previously introduced. We choose such a size of the tracer population to guarantee the convergence of the statistical analyses, while still preserving a reasonably short computational time.
The tracers motion is determined by the $\mathbf{E} \times \mathbf{B}$ drift velocity, i.e. we reconstruct the trajectory of each tracer from the numerical integration of the first-order dynamical system 
\begin{align}\label{dymsys1}
    &\dot{u}= -\partial_v \Phi\\\label{dymsys2}
    &\dot{v}= \partial_u \Phi,
\end{align}
where the overdot indicates derivation with respect to the dimensionless time $\tau$. The time integration is carried out following the same prescriptions implemented in the case of the electric potential. 

Due to the turbulent nature of the electrostatic potential $\Phi$, the dynamical system \eqref{dymsys1}-\eqref{dymsys2} can be seen as a couple of Langevin equations. Therefore, we consider the displacements with respect to the initial position of each tracer at a given time, namely $\Delta u(\tau)=u(\tau)-u(0)$ and $\Delta v(\tau)=v(\tau)-v(0)$, as a couple of random variables on which we perform statistical analyses in order to study the underlying 2D distribution function. Specifically, by computing the second moments $\sigma_{uu}(\tau)=\left < \leri{\Delta u(\tau)}^2 \right>$ and $\sigma_{vv}(\tau)=\left < \leri{\Delta v(\tau)}^2 \right>$, where $\left < \cdot \right >$ represents an averaging over the ensemble of tracers, it is possible to calculate the mean squared displacement (MSD) as a function of time via the formula $\text{MSD}(\tau)=\sigma_{uu}(\tau)+\sigma_{vv}(\tau)$. This quantity describes the different regimes of transport to which the tracers population is subject in any phase of the simulation. Indeed, the time dependence of the MSD can be parameterized as 
\begin{equation}
    \text{MSD}(\tau)\propto \tau^\gamma
\end{equation}
and a value of the exponent $\gamma$ greater (smaller) than $1$ indicates a supradiffusive (subdiffusive) behavior, whereas the diffusive regime corresponds to the $\gamma=1$ case. In other words, it is possible to calculate the turbulent transport coefficient $\mathcal{D}_T(\tau)$ as 
\begin{equation}
\mathcal{D}_T(\tau)=\frac{\text{MSD}(\tau)}{4\tau}.
\end{equation}
A standard diffusive phase is then connected to a linear-in-time behavior of MSD, thus generating a constant $\mathcal{D}_T$. We remark that the quantity $\mathcal{D}_T$ has not a direct connection with the diffusion coefficient and the viscosity appearing in the dynamical equations of our model, namely the parameters $\mathcal{D}$ and $\nu$ respectively. The latters play a significant role when the the calculation of the electric potential is performed via the integration of the system    \eqref{eqn}, \eqref{eqvort}, \eqref{eqohm} (or merely of equation \eqref{tordued} if the reduced model for pure toroidal field is considered). In this work we are interested in studying the transport properties of the electric potential itself due to its turbulent nature. To do so we resort to the calculation of the MSD, which has been linked \cite{https://doi.org/10.1112/plms/s2-20.1.196,Batchelor_1952} to the eddy diffusivity tensor entering the Reynolds-averaged version of Navier-Stokes equation \cite{doi:10.1098/rsta.1895.0004}. In section \ref{seca1} we will a posteriori verify that the turbulent diffusion coefficient obtained from the passive tracers trajectories analysis is indeed the same object (or a very good approximation of it) involved in one of the closure schemes of the averaged Navier-Stokes equation, as for instance the $K-\epsilon$ model \cite{JONES1972301}. 

Since the tracers tend to exit from the simulation region following the advective $\mathbf{E}\times\mathbf{B}$ field we adopt the following prescription: we dub as statistically significant only the portion of tracers which remain at every time of our simulation within a square, with side length equal to $20\pi$, centered in the origin of our 2D coordinate system. In this manner we are able to correctly compute the MSD in unwrapped coordinates \cite{QIAN1991910} and obtain meaningful physical quantities.

Apart from the study of turbulent transport with the methods described above, we perform normality tests on the 2D distribution function of the random variables $\Delta u(\tau)$ and $\Delta v(\tau)$, quantifying its departure from Gaussianity. The symmetry with respect to the central value is investigated through the computation of a vector-valued measure of skewness, as given in \cite{c2aa02eb-fd22-3300-8d14-27844fc8cabc,article}. This asymmetry estimator is constructed in the following manner: given a 2D random vector $\mathbf{x}=\leri{x_1,x_2}^T$ with mean $\bm{\mu}$, non-singular covariance matrix $\Sigma$ with elements
\begin{equation}
    \Sigma= \begin{pmatrix}
        \sigma_{1,1} &\sigma_{1,2}\\
        \sigma_{1,2} & \sigma_{2,2}
    \end{pmatrix}
\end{equation}
and finite third order moments, it is possible to define the standardized vector $\mathbf{z}=\leri{z_1,z_2}^T=\Sigma^{-\frac{1}{2}}\leri{\mathbf{x}-\bm{\mu}}$, having by construction the null vector as mean value and the two-dimensional identity as covariance matrix. The vector-valued skewness estimator here adopted is a function $\mathbf{s}\leri{\mathbf{x}}: \mathbb{R}^2 \to \mathbb{R}^2$, such that
\begin{equation}
    \mathbf{s}\leri{\mathbf{x}}=\mathbf{s}\begin{pmatrix}
        x_1 \\
        x_2
    \end{pmatrix}= \begin{pmatrix}
        \left < z_1^3\right >+ 
\left < z_1^2z_2\right > \\
         \left <z_2^3\right >+ \left <z_2^2z_1\right >
    \end{pmatrix}.
\end{equation}
 Each component of the vector $\mathbf{s}=\leri{s_1,s_2}^T$  accounts for an asymmetry estimator on the relative direction. Specifically, a positive (negative) sign of a component of $\mathbf{s}$ indicates that the mean value is greater (smaller) than the mode on the corresponding direction, whereas a null component signals that the section of the 2D distribution function along that specific direction results symmetrical.

Another normality test we perform is the assessment of the outliers abundance, i.e. the weight of the distribution tales with respect to a reference Gaussian with equal variance. This task is carried out through the computation of a scalar-valued measure of kurtosis, i.e. a function $\mathcal{K}:\mathbb{R}^2\to\mathbb{R} $, whose explicit expression reads 
\begin{multline}
    \mathcal{K}= \frac{1}{\leri{\sigma_{1,1}\sigma_{2,2}-\sigma_{1,2}^2}^2} \bigg [ \sigma_{1,1}^2 \left < x_2^4\right >+\sigma_{2,2}^2 \left < x_1^4\right >+2\sigma_{1,1}\sigma_{2,2} \left < x_1^2x_2^2\right > -\\
     -4 \sigma_{1,2}\leri{\sigma_{1,1}\left < x_1x_2^3\right >+\sigma_{2,2}\left < x_1^3x_2\right >-\sigma_{1,2}\left < x_1^2x_2^2\right >} \bigg].
\end{multline}
This quantity is equal to $8$ in the case of a 2D normal distribution and values greater (smaller) than $8$ denote heavier (lighter) tails. Specifically, we speak either of leptokurtic or platykurtic curves referring to distributions with excess or deficiency of kurtosis, respectively.

\section{Results}\label{sec4}
Here we provide an overview of the results achieved from the study of the fluid passive tracers motion along the advective velocity given by the $\mathbf{E} \times \mathbf{B}$ drift. The analysis has been carried out in three distinct scenarios: 
\begin{itemize}
    \item[A1 -] the ideal inviscid case with a pure toroidal magnetic field, i.e. the electric potential is obtained from the integration of \eqref{tordued} with $\alpha_2=0$;
    \item[A2 -] the viscous case, still with a toroidal magnetic field, i.e. same as in case A1 but setting $\alpha_2 \neq 0$;
    \item[B\phantom{1} -] assuming a background X-point magnetic configuration with a non-null viscosity coefficient, i.e. the electric potential is obtained from the integration of the system \eqref{eqn}, \eqref{eqvort}, \eqref{eqohm}.
\end{itemize}
For each of the cases described above, we present the outcome of two different scenarios, namely the electric field on which we evolve the tracers either undergo the process of condensation or the latter is absent from the field evolution at any time. Our goal is therefore to investigate the role of three distinct physical features, namely viscosity, magnetic field geometry and condensation, as well as possible interactions between them. In each section we describe the outcome of our analyses, providing a description of turbulent transport via the descriptive tools introduced in the previous section, namely the MSD and the instantaneous transport coefficient. In addition to this, we show the result of the normality tests performed on the tracers distribution function.

\subsection{Case A1}\label{seca1}

   In this section we start to present the results obtained from our analysis of the turbulent transport induced by equation \eqref{tordued}, hence in the absence of effects due to the X-point magnetic configuration, on fluid passive tracers. The first scenario we consider is the ideal inviscid setting, obtained by imposing $\alpha_2=0$. 
   Rather than providing a description of a realistic physical system, the analysis performed in the ideal case constitutes a benchmark against which a comparison with the viscous version of \eqref{tordued} can be carried out, in order to precisely assess the effect of viscosity on the transport phenomenon. 
   \begin{figure}[hbtp!]
  \begin{subfigure}[c]{.48\linewidth}
    \centering
    \includegraphics[width=\linewidth]{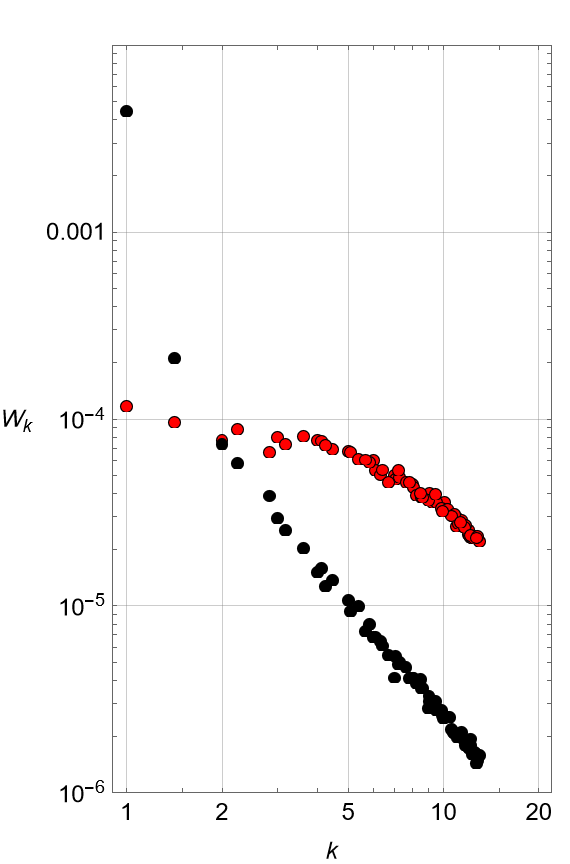}%
  \end{subfigure}\hfill
  \begin{tabular}[c]{@{}c@{}}
    \begin{subfigure}[c]{.48\linewidth}
      \centering
      \includegraphics[width=0.7\linewidth]{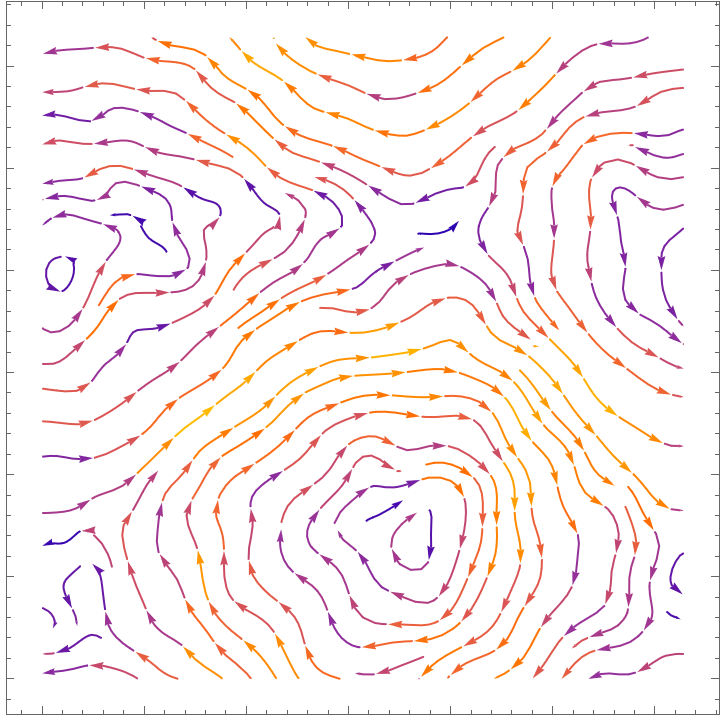}
    \end{subfigure}\\
    \noalign{\bigskip}%
    \begin{subfigure}[c]{.48\linewidth}
      \centering
      \includegraphics[width=0.7\linewidth,page=2]{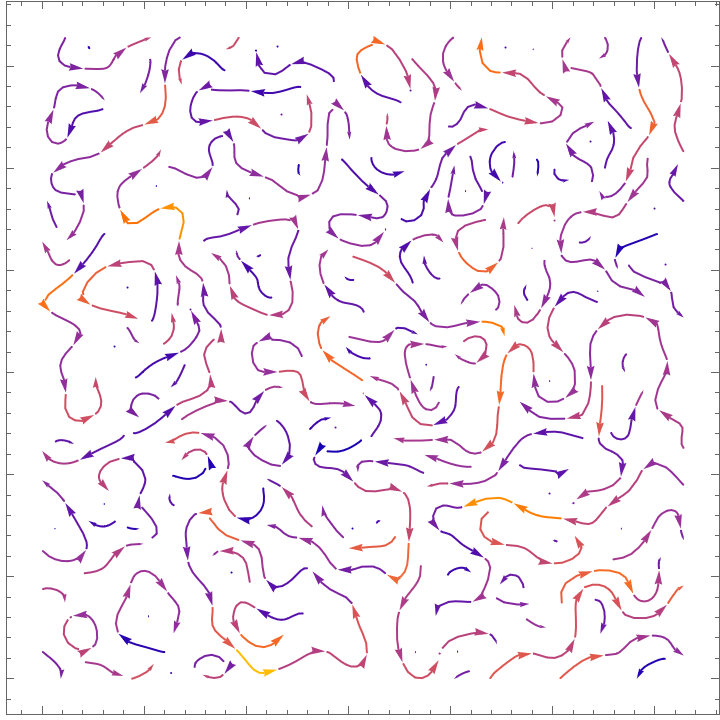}
    \end{subfigure}
  \end{tabular}
  \caption{Case A1. In the left panel we report the energy spectrum, indicating with black and red dots the condensation and non-condensation case, respectively. In the right panels we show the advective $\mathbf{E}\times \mathbf{B}$ field in the poloidal plane $(u,v)$ for the condensation (top) and non-condensation case (bottom). Both the energy spectrum and the advective fields are taken at the end of the simulation time, namely for $\tau=9500$.}
      \label{fig:every}
  \end{figure}
   In Fig. \ref{fig:every} we show, for both the condensation and non-condensation cases, the energy spectrum and the advective $\mathbf{E}\times\mathbf{B}$ field obtained at the end of the simulation time.
   As previously stated, we can notice that, in the condensation case, the energy of the system is properly stored in the first few modes, while the remaining part of the spectrum follows a profile $\propto k^{-2}$, as predicted by the theory \cite{10.1063/1.1762301,Kraichnan_1975,10.1063/1.861243}. For what concerns the evolution of the system in the absence of condensation, instead, a quasi-flat spectrum emerges, signaling that eddies of different sizes roughly share the same amount of energy. 
   
\begin{figure}[hbtp!]
\centering
\includegraphics[width=6cm]{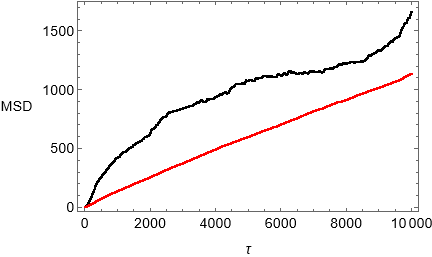}\!\!
\includegraphics[width=6cm]{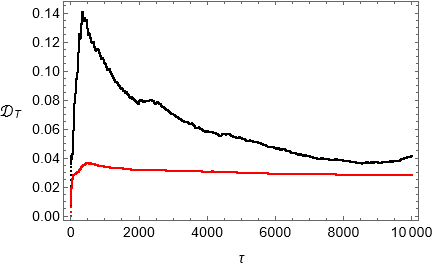}
\caption{Case A1. The mean squared displacement and the turbulent transport coefficient are plotted in the left and right panel, respectively. The black curves describe the condensation case, non-condensation in red.}
\label{fig:msddiffnv}
\end{figure}

\begin{figure}[hbtp!]
\centering
\includegraphics[width=5cm]{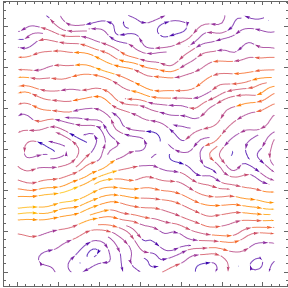}
\!\!\!
\includegraphics[width=5cm]{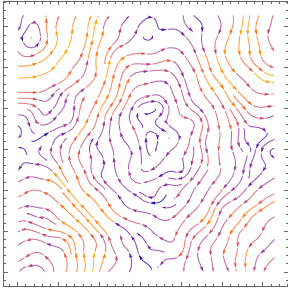}
\caption{Case A1. The $\mathbf{E}\times \mathbf{B}$ field at $\tau=300$ (left panel) and $\tau=1300$ (right panel) in the condensation scenario.}
\label{fig:openclosed}
\end{figure}

   Let us begin the tracer analysis by presenting the MSD, quantifying the mean departure from the initial position and, in a broad sense, the magnitude of transport, in the left panel of Fig. \ref{fig:msddiffnv}. The first aspect calling for a comment is the greater amount of transport observable in the condensation case. Indeed, as it can be noticed from the plots, the MSD described by the black curve is roughly $50 \%$ larger than the one in red at the end of the simulation time.
Moreover, it can be easily noticed that condensation causes the MSD to undergo different phases characterized by subdiffusive and supradiffusive behaviors, whereas in the absence of condensation the MSD grows linearly in
time. 
This aspect can be better emphasized by directly computing the instantaneous transport coefficient, resulting in the plots presented in the right panel of Fig. \ref{fig:msddiffnv}. By looking at the red curve, we can observe that the evolution of the system obtained by initializing the electric potential integration so that condensation is prevented is basically a quasi-pure diffusive process: indeed, after a short transient initial phase of supradiffusive transport, $\mathcal{D}_T$ stabilizes around a value $\simeq 0.030$ roughly for $\tau=1500$ until the end of the simulation time. The same does not hold for the condensation case: as it can be noticed from the black curve, the system undergo a series of different transport regimes. Specifically, we outline the presence of a first phase between $\tau=0$ and $\tau \simeq 1000$ in which the transport coefficient rapidly grows with time, signaling a supradiffusive behavior associated with advective transport. Then, from $\tau\simeq 1000$ until $\tau\simeq 2000$ we observe a subdiffusive regime. These two early stages of the system evolution can be explained by a direct inspection of the $\mathbf{E}\times \mathbf{B}$ field, of which we display snapshots taken at $\tau=300$ and $\tau=1300$ in the left and right panel of Fig. \ref{fig:openclosed}, respectively. 

\begin{figure}[hbtp!]
\centering
\includegraphics[width=6cm]{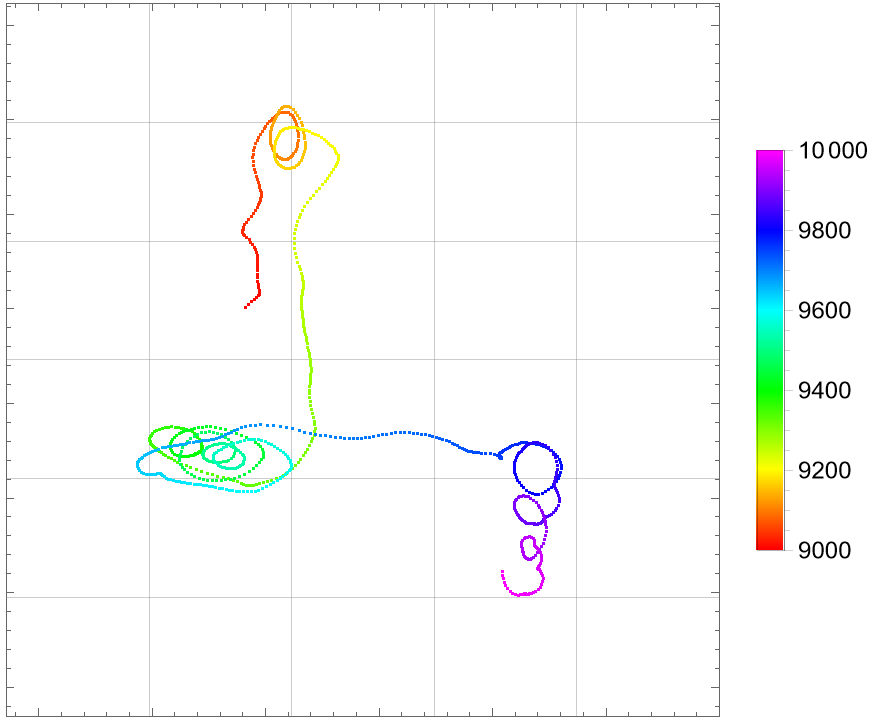}\!\!\!\!\!\!\!\!\!\!\!\!\!\!\!\!\!\!\!
\includegraphics[width=6cm]{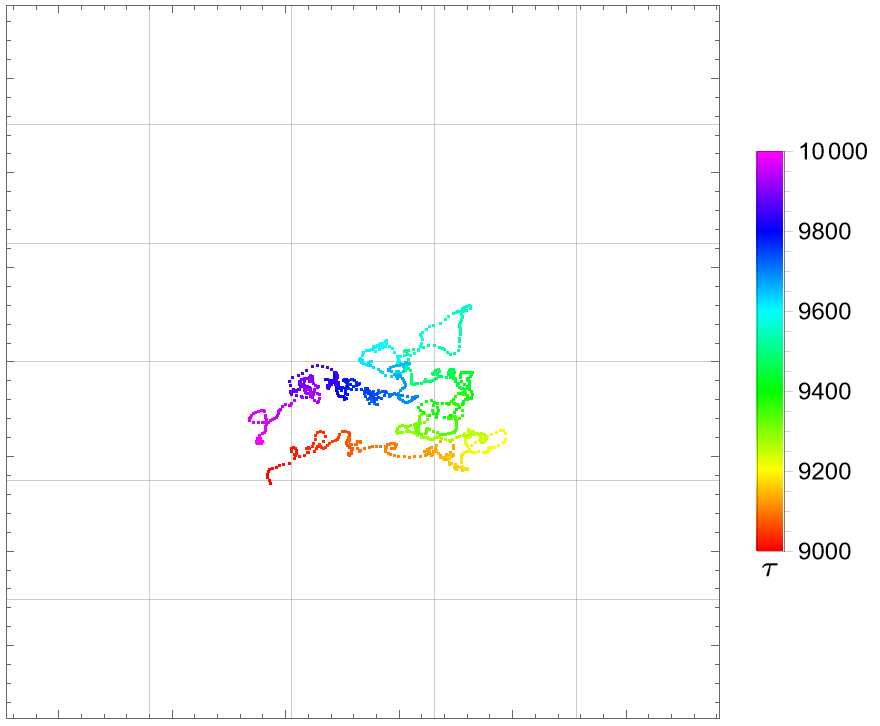}
\caption{Case A1. Portion of the $(u,v)$ poloidal plane with the trajectory of a single tracer, for the condensation (left panel) and non-condensation cases (right panel) starting in $\tau=9000$ (red) and ending in $\tau=10000$ (purple).}
\label{fig:trajCNV}
\end{figure}

As it can be noticed from the plot in the left panel, at $\tau=300$ most of the streamlines are open curves and this results in advective transport for the vast majority of the tracer population. By looking instead at the $\mathbf{E}\times \mathbf{B}$ field at $\tau=1300$ (right panel) we notice that the overall morphology of the streamlines is deeply changed. Indeed, in this case we see that most of the tracer trajectories are along closed curves, so that a great portion of tracers are trapped in a vortex, resulting in a greater confinement for the whole population. Following these early transients,
 we can observe that there are two distinct diffusive phases: a first short one between $\tau \simeq 2000$ and $\tau \simeq 2700$ characterized by a diffusion coefficient $\simeq 0.08$ followed by a long subdiffusive relaxation towards a second asymptotic diffusive regime with $\mathcal{D}_T\simeq 0.04$. Having this in mind, we can conclude that, for what concerns the inviscid dynamics in the case of a pure toroidal magnetic field, condensation acts as an enhancer of transport, with a maximum value of the instantaneous transport coefficient $\simeq 0.14$, in contrast to the correspondent maximum obtained in the absence of condensation amounting to $\simeq 0.035$. In order to give an explanation of this fact, we have to recall the microscopic interpretation of a typical diffusive process, i.e. a random walk caused by stochastic impacts. In the context of turbulent diffusion a single particle gets scattered every time it interacts with a vortex and the corresponding mean free path is connected to the typical distance between vortices. As it can be noticed from Fig. \ref{fig:trajCNV}, the trajectory of a tracer in the late stages of the simulation is radically different when the condensation and non-condensation cases are compared. The large eddies size characterizing the condensation case is responsible for a greater transport, given that a single tracer is typically trapped for a while inside a vortex, then gets expelled from it and start roaming almost unperturbed until the next encounter with another vortex. The larger amount of transport resulting on average for the whole tracer population is therefore due to the greater distance between eddies. 
On the contrary, when the non-condensation case is addressed, we notice that a single tracer gets scattered much more often or, in other words, the free path between encounters with vortices is far smaller. 
Moreover, in the non-condensation scenario, the interval of time in which a tracer gets trapped inside a vortex is basically null and the obtained trajectory resembles much more a proper random walk, hence explaining the quasi-pure diffusive character of transport observed.

It is now important to point out that the transport coefficients calculated above are not related to the diffusion and viscosity coefficients appearing in the equation of the model, as previously stated in section \ref{secsetup}. Indeed, the object $\mathcal{D}_T$ here obtained from the tracers motion analysis quantifies the rate of transport due to turbulence only, and must be therefore linked to the turbulent diffusivity tensor emerging from the averaging of Navier-Stokes equation (to which the basic equation \eqref{tordued} of our reduced model is equivalent). The expression of the eddies diffusivity tensor in terms of mean flow quantities is the problem addressed in the closure schemes necessary to complete the fluid averaged theory. In this work we will consider, as theoretical paradigm, the $K - \epsilon$ model, from which it is possible to calculate the turbulent coefficient for pure diffusive regimes as 
\begin{equation}
    \mathcal{D}_T=c_\mu \mathcal{N} \sqrt{K} L.
\end{equation}
Here $c_\mu$ is a constant characterizing the model and its value is empirically determined to be $c_\mu\simeq 0.09$, while $L$ is the typical size of eddies in the physical system under consideration. The latter can be deduced from the data reported above as the square root of the MSD taken at the beginning of the diffusive phase \cite{2023JPlPh..89a9008S}.
For instance, in the condensation case we have seen that a short diffusive phase begins around $\tau\simeq 2000$ and the corresponding value of the MSD amounts to roughly $610$, hence $L\simeq 24.7$.
The kinetic energy $\bar{K}$ can be calculated from the initial Fourier mode and the relative amplitudes, from \eqref{costantimoto}, resulting in $\bar{K}\simeq 1.4\times 10^{-3}$. We thus obtain $\mathcal{D}_T\simeq 0.082$, reproducing with great precision the value deduced from the data analysis. Following the same steps we calculate the turbulent diffusion coefficient in the non-condensation case as $\mathcal{D}_T\simeq 0.049$. Here the discrepancy between theory and simulation slightly grows with respect to the previous case, but we can nevertheless affirm that the parameter $\mathcal{D}_T$ calculated from the tracers analysis satisfactorily approximates (at least catching the right order of magnitude) the object predicted by the closure model.

\begin{figure}[hbtp!]
\centering
\includegraphics[width=7cm]{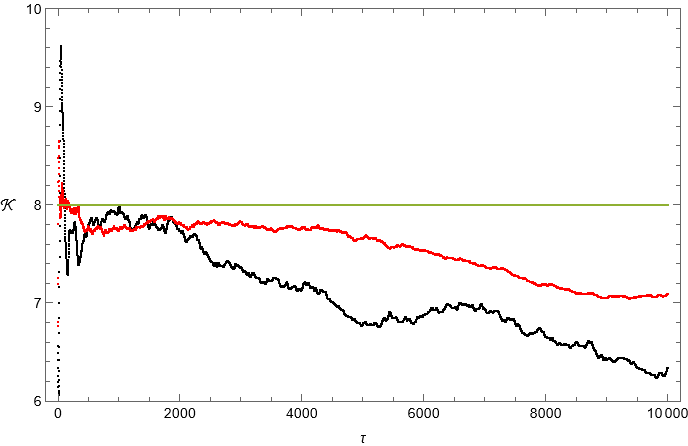}
\caption{Case A1. Kurtosis: the condensation case is described by the black curve, non-condensation in red. The green line indicates the reference value for a Gaussian distribution.}
\label{fig:kurtnv}
\end{figure}

We now focus on the normality tests, performed in the spirit of highlighting departures of the 2D probability distribution from a Gaussian profile. 
\begin{figure}[hbtp!]
\centering
\includegraphics[width=6cm]{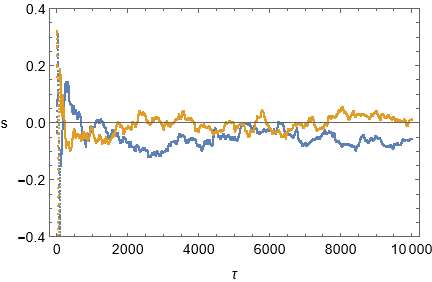}\!\!
\includegraphics[width=6cm]{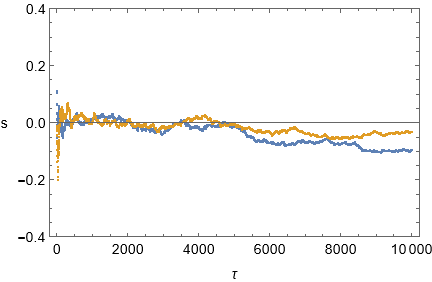}
\caption{Case A1. Skewness: condensation (left) and non-condensation (right). The blue curve describes the skewness along the $u$ direction, orange color for the $v$ direction.}
\label{fig:skewnv}
\end{figure}
As previously stated, we calculate estimators of the third and fourth moments of the tracers position distribution, in order to quantify skewness and kurtosis, whose behaviors are reported in Fig. \ref{fig:skewnv} and Fig. \ref{fig:kurtnv}, respectively. The symmetry test conducted through the computation of skewness shows that the deviation from a Gaussian profile is rather small. The precise outcome of this test can randomly vary, within a narrow range around $0$, depending on the specific choice of initialization values for the field integration. In spite of this, we claim that, with great generality, the skewness test returns values sufficiently close to $0$ to affirm that the 2D distribution of the tracers position is characterized by a symmetric shape.

The kurtosis test, instead, shows a tendency which remains stable with respect to changes in the initialization values of the field integration. Indeed, as it can be clearly seen from Fig. \ref{fig:kurtnv}, both in the presence and in the absence of condensation, a significant departure from a Gaussian profile is highlighted.
 Specifically, in both cases we observe that the scalar-valued estimator of kurtosis adopted in this work takes a value $\simeq 7$, pointing out a smaller amount of outliers with respect to a Gaussian distribution having the same variance. This is certainly not surprising for the condensation scenario: as we shown through the analysis of both the MSD and the instantaneous transport coefficient, in the condensation case many different regimes of diffusive and anomalous transport characterize the system evolution, the latter being far from a pure diffusive process. It is therefore natural to observe a departure from Gaussianity also in the case of the kurtosis estimator. For what concerns the non condensation scenario, the issue is certainly more subtle and deserves further investigations. However, we remark that the departure from a pure Gaussian behavior is in this case rather small (basically we are looking at a $10\%$ deviation) and a possible explanation for this anomaly includes some sort of residual trapping experienced by the tracers.
 
\subsection{Case A2}\label{seca2}
\begin{figure}[hbtp!]
  \begin{subfigure}[c]{.48\linewidth}
    \centering
    \includegraphics[width=\linewidth]{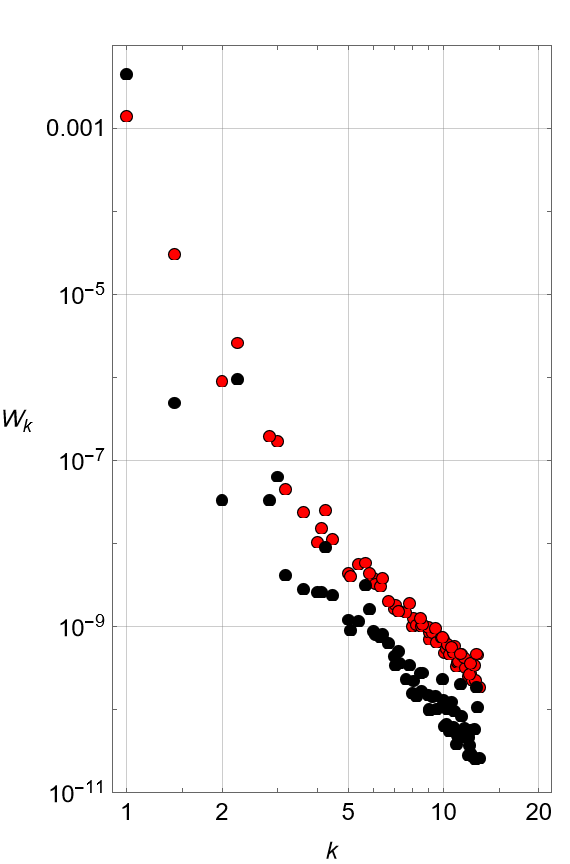}%
  \end{subfigure}\hfill
  \begin{tabular}[c]{@{}c@{}}
    \begin{subfigure}[c]{.48\linewidth}
      \centering
      \includegraphics[width=0.7\linewidth]{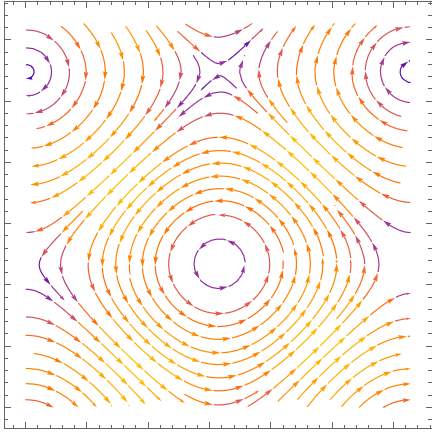}
    \end{subfigure}\\
    \noalign{\bigskip}%
    \begin{subfigure}[c]{.48\linewidth}
      \centering
      \includegraphics[width=0.7\linewidth,page=2]{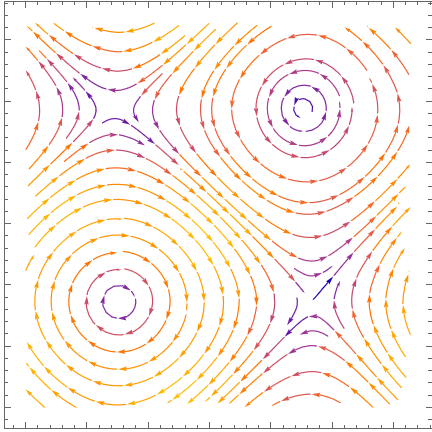}
    \end{subfigure}
  \end{tabular}
  \caption{Case A2. Left panel: energy spectrum for the condensation (black) and non-condensation (red) cases. Right panels: the advective $\mathbf{E}\times \mathbf{B}$ field in the poloidal plane $(u,v)$ for the condensation (top) and non-condensation case (bottom). Both plots refer to the end of the simulation time, namely for $\tau=9500$.}
      \label{fig:everyvisc}
  \end{figure}

We now proceed to present the results obtained from our analysis of transport induced by equation \eqref{tordued} but, contrarily to the previous section, here we consider the viscous case by setting  $\alpha_2\neq 0$. 
As before, we start by presenting the spectrum and the advective $\mathbf{E}\times\mathbf{B}$ field, in Fig. \ref{fig:everyvisc}.
\begin{figure}[hbtp]
\centering
\includegraphics[width=6cm]{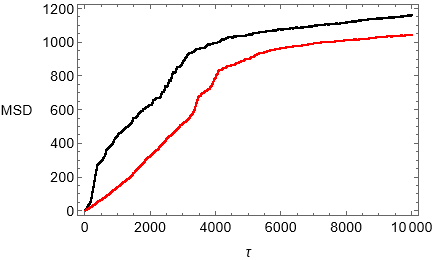}\!\!
\includegraphics[width=6cm]{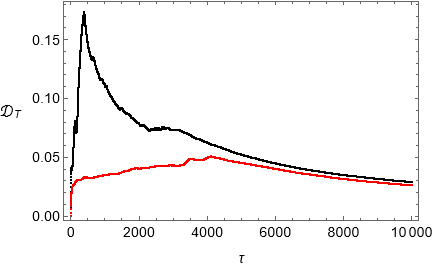}
\caption{Case A2. MSD and $\mathcal{D}_T$ for the condensation (black) and non-condensation (red) cases.}
\label{fig:msddiffv}
\end{figure}
By comparing the plots here reported with their analog in Fig. \ref{fig:every} we notice that the introduction of viscosity causes the dominance of large scale structures in the final stages of the system evolution, even in the absence of condensation. Hence, we expect that, from a certain point of our simulation time, the condensation and non-condensation cases should roughly overlap, in contrast to the ideal case presented in the previous section. 

We begin to display the results obtained from the tracer analysis by providing plots of the MSD behavior in time, as given in the left panel of Fig. \ref{fig:msddiffv}. We can immediately observe that, either in the presence or absence of condensation, the total amount of transport has lowered with respect to the inviscid case. This tendency has a more profound impact in the presence of condensation: indeed, in this case, the introduction of viscosity is responsible for a reduction of the MSD at the end of the simulation time of roughly $25 \%$. For what concerns the simulation in the absence of condensation we notice instead a decrease of the MSD maximum value of about $10 \%$. The introduction of viscosity plays a major role also in modifying the overall shape of the curves: indeed, while in the inviscid case the MSD profiles are qualitatively different from each other, we can see that they acquire a similar behavior when a non-null viscosity is taken into account. It is the curve describing the non-condensation case that results particularly modified: from the observation of the MSD profile we can affirm that a number of phases of anomalous transport are present during the system evolution, in contrast with the inviscid case in which the MSD grows as a straight line for almost the entire simulation time.
\begin{figure}[hbtp!]
\centering
\includegraphics[width=6cm]{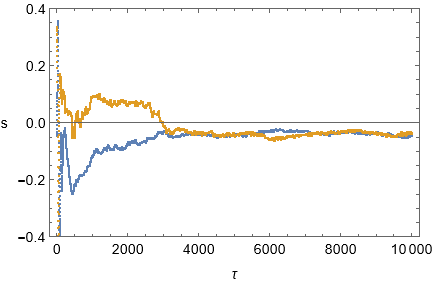}\!\!
\includegraphics[width=6cm]{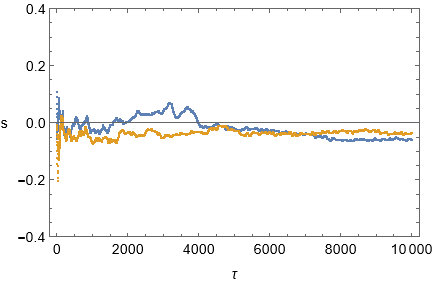}
\caption{Case A2. Components of the skewness estimator vector for the condensation (left) and non-condensation (right) cases. The blue (orange) curve describes the skewness along the $u$ ($v$) direction.}
\label{fig:skewcv}
\end{figure}
This fact can be further appreciated from a direct calculation of the instantaneous transport coefficient, reported in the right panel of Fig. \ref{fig:msddiffv}. Let us focus on the red curve describing the behavior in time of the transport coefficient in the absence of condensation: the qualitative discrepancy of this curve with the analog obtained in the inviscid case is significant. Indeed, while in the previous section we noticed a diffusive transport throughout all stages of the system evolution, here we see that a well-defined diffusive phase is almost not recognizable. More precisely, the instantaneous transport coefficient has practically the same behavior as in the inviscid for very short times, i.e. for $\tau$ between $0$ and $500$: a steep ramp signaling a highly supradiffusive phase followed by a plateau whose value results around $0.03$. This similarity can be easily explained given that, on such short time scale, the effect of viscosity is obviously negligible. However, for times greater than $\tau=500$ we see that the introduction of the viscous term in \eqref{tordued} causes a completely different behavior of the instantaneous transport coefficient. Indeed, we observe that a supradiffusive regime characterizes the system until $\tau \simeq 4000$, followed by a subdiffusive phase leading to an asymptotic diffusive regime, matching the behavior of the condensation case for late times. The curve describing the latter, in black, overlaps very closely its analog obtained in the absence of viscosity. We can provide a simple explanation of this fact by observing that condensation and viscosity qualitatively act in the same manner on the system morphology: both these physical features are responsible for a dominance of the large spatial scales on the system energy distribution, leading to a landscape of large eddies prevailing from a certain stage of the time evolution until its end (as pointed out by the plots in Fig. \ref{fig:every}). This fact has also an impact on the transport coefficient calculated in the absence of condensation. In facts we can observe that the maximum value of the transport coefficient is slightly greater in the viscous case with respect to inviscid one. This observation confirms that big vortices are, in general, responsible for an enhancement of transport. However, the effect of viscosity is also to lower the total energy of the electric field, hence leading to a minor transport for late stages of the evolution, as signaled by the decrease of the maximum value of the MSD in both the condensation and the non-condensation cases. 

The comparison of the tracer data with the $K - \epsilon$ model will not be repeated for the viscous case presented in this section. Indeed, by comparing Fig. \ref{fig:msddiffnv} and Fig. \ref{fig:msddiffv} we see that both in the presence and in the absence of condensation, the two early diffusive phases are found at roughly the same times with respect to their analog in the inviscid case. Therefore, the estimate of the parameter $L$ entering the definition of the turbulent diffusion coefficient gives approximately the same result as in the previous section. In conclusion, the theoretical background offered by the $K - \epsilon$ model is, also in this case, well-grounded and able to predict with adequate precision the numerical data coming from the tracer analysis. 

\begin{figure}[hbtp!]
\centering
\includegraphics[width=7cm]{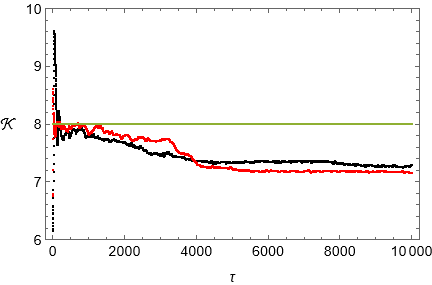}
\caption{Case A2. The scalar estimator of kurtosis $\mathcal{K}$ as a function of time. The black (red) curve refers to the condensation (non-condensation) case. The green line indicates the reference value for a Gaussian distribution.}
\label{fig:kurtcv}
\end{figure}

The introduction of viscosity has a far less significant effect on the normality of the distribution function describing the tracers positions. As we can observe from Fig. \ref{fig:skewcv} and Fig. \ref{fig:kurtcv}, the behavior of skewness and kurtosis greatly overlaps their analog in the inviscid case. Indeed, as seen in the previous section, the asymmetry of the distribution remains sufficiently close to $0$, whereas a slight deviation from a genuine Gaussian behavior concerns the number of outliers populating the tails of the distribution, the latter resulting smaller than the standard values implied by normality.

We conclude this section by remarking that, contrarily to the ideal case displayed above, here we do not present the single tracer trajectory for late times. This is because, as observed from Fig. \ref{fig:everyvisc}, the introduction of viscosity makes the condensation and non-condensation cases rather similar in the final part of the system evolution, as already highlighted. 

\subsection{Case B: Effect of the X-point magnetic configuration}
\begin{figure}[hbtp!]
  \begin{subfigure}[c]{.48\linewidth}
    \centering
    \includegraphics[width=\linewidth]{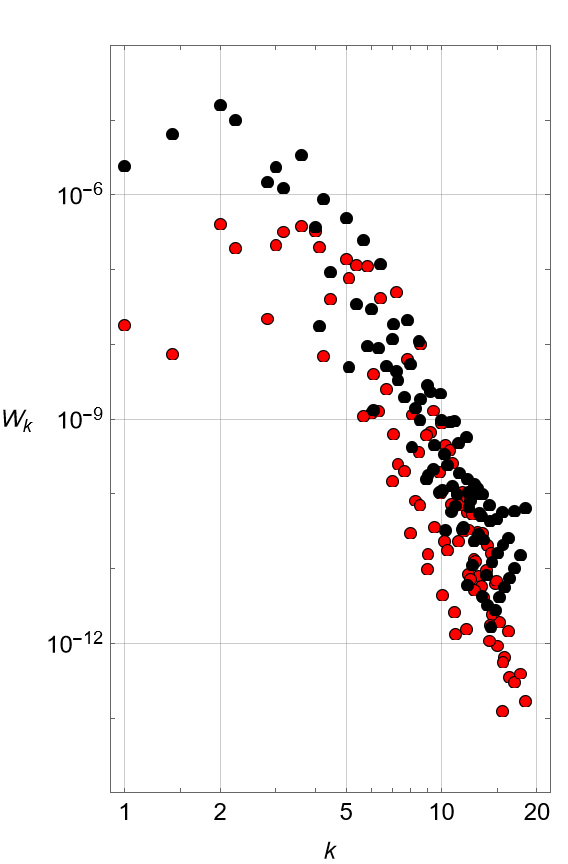}%
  \end{subfigure}\hfill
  \begin{tabular}[c]{@{}c@{}}
    \begin{subfigure}[c]{.48\linewidth}
      \centering
      \includegraphics[width=0.7\linewidth]{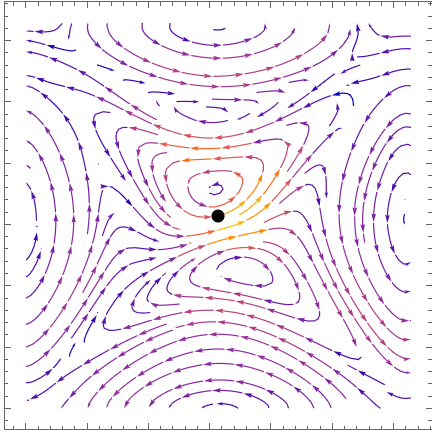}
    \end{subfigure}\\
    \noalign{\bigskip}%
    \begin{subfigure}[c]{.48\linewidth}
      \centering
      \includegraphics[width=0.7\linewidth,page=2]{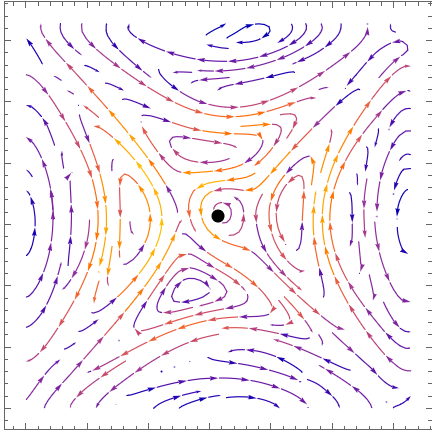}
    \end{subfigure}
  \end{tabular}
  \caption{Case B. Left panel: energy spectrum for the condensation (black) and non-condensation (red) cases. Right panels: the advective $\mathbf{E}\times \mathbf{B}$ for the condensation (top) and non-condensation case (bottom). Both the plots refer to the end of the simulation time, namely for $\tau=9500$. The black dot indicates the location of the X-point.}
      \label{fig:everyxp}
  \end{figure}

Having presented the outcome of our simulations in the case of a toroidal magnetic field, both in the absence and in the presence of viscosity, we now proceed to display the results obtained when the X-point magnetic geometry is taken into account. To do so, we resort to the complete set of equations governing the full model as given in \eqref{eqn}, \eqref{eqvort} and \eqref{eqohm}. We remark that here we will only consider the viscous case in order to faithfully describe a realistic scenario, having elucidated in the previous sections what are the main changes occurring when viscosity is turned on starting from an ideal inviscid setting. In Fig. \ref{fig:everyxp} we display the spectrum and the advective field obtained in this case. As it can be noticed, the condensation and non-condensation cases are rather similar in the final stages of the integration, due to a non-null viscous term in the equations considered. Hence, as in the previous section, we expect that the tracer dynamics be roughly the same in the two cases, at least in the asymptotic regime.
  
\begin{figure}[hbtp!]
\centering
\includegraphics[width=6cm]{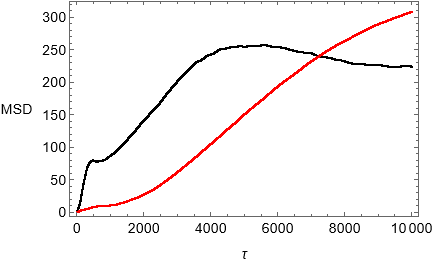}\!\!
\includegraphics[width=6cm]{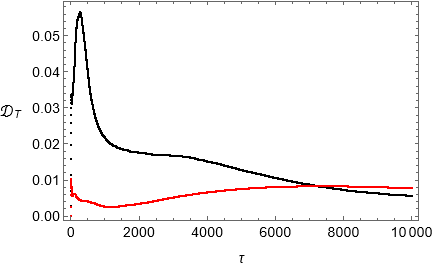}
\caption{Case B. The mean squared displacement and the turbulent transport coefficient are plotted in the left and right panel, respectively. The black curves describe the condensation case, non-condensation in red.}
\label{fig:msddiffpx}
\end{figure}

Let us begin by analyzing the MSD, depicted in the plots reported in the left panel of Fig. \ref{fig:msddiffpx}.  
Several features of these curves are radically different from the ones presented in the previous sections. For instance, here the total amount of transport, deduced from the MSD maximum value, is in this case larger in the absence of condensation (red curve). Comparing the curves in black and red, it is easy to see that transport is lowered by an amount of roughly the $20 \%$ when condensation is present (black curve). Moreover, while the MSD calculated in the absence of condensation monotonically increases with time, as every other MSD curve presented above, the function representing the MSD in the condensation case presents a maximum for a time $\tau\simeq 5000$, then slightly decreases in the subsequent phase, relaxing to a constant plateau in the final stage of the simulation. The value of the MSD at the plateau is about $10 \%$ smaller than the maximum reached at mid-run. In addition to this, it is important to notice that the introduction of the X-point magnetic configuration acts as a strong limiter of particle transport. By comparing the left panels of Fig. \ref{fig:msddiffv} and Fig. \ref{fig:msddiffpx} we observe that the maximum MSD is reduced by a factor of roughly $1/5$ and $1/3$ in the condensation and non-condensation cases, respectively. This fact can be easily explained by noticing that the X-point divides the square constituting our integration domain into four (quasi) disconnected regions of triangular shape and the presence of a magnetic shear significantly reduces turbulent transport among them. Hence, the fraction of tracers initialized in one specific region that migrates in a different one is very low. It is straightforward to recognize this as the ultimate cause of the mitigated transport when the X-point magnetic configuration is taken into account. We remark that the adopted background magnetic geometry reduces the maximum length scale of vortices, as it can be deduced from the energy spectrum in Fig. \ref{fig:everyxp}. Indeed, in this case we see that the maximum energy is stored in the $k=2$ mode, for both the condensation and non-condensation scenario. This feature can be explained by noticing that the existence of vortices having size roughly equal to the size of the integration domain (associated to the wavenumber $k=1$) is prevented by the introduction of the X-point magnetic field, since these large scale structures would necessarily intersect the separatrix.

We now turn to the transport coefficient, whose plots are given in the right panel of Fig. \ref{fig:msddiffpx}.
Let us focus on the black curve describing the condensation case: by comparing this profile with the analog in Fig. \ref{fig:msddiffv} we see that the introduction of the X-point does not significantly alter the transport coefficient behavior at a qualitative level. Indeed, we can still recognize two very short phases, namely a supradiffusive followed by a subdiffusive regimes, leading to pure diffusive transport between $\tau\simeq 1750$ and $\tau\simeq 3500$. Then, from around half of the simulation time until the end, we observe a slow subdiffusive ramp ending in an asymptotic diffusive regime. At a quantitative level instead, we see that the presence of the X-point causes a reduction of the transport coefficient value by a factor between $1/3$ and $1/4$, depending on the particular phase considered. Let us now focus on the non-condensation case: by comparing the red curve reported in the right panel of Fig. \ref{fig:msddiffpx} with its analog plotted in Fig. \ref{fig:msddiffv}, we observe profound changes both in the overall morphology of the curves and in the values reached by the transport coefficient. A first comment is that, beside a very short diffusive phase between $\tau\simeq 1500$ and $\tau\simeq 1700$, a quasi-constant plateau appears between $\tau\simeq 6000$ and $\tau\simeq 10000$. Furthermore, the concavity of the curve for late times does not signal the arising of a relaxation towards an asymptotic diffusive regime. For what concerns the values of the transport coefficient we highlight a reduction even more severe that the one noticed in the condensation case. Indeed, for the curve obtained in the absence of condensation we measure a transport coefficient roughly $6$ times smaller than its analog calculated in the slab case. This being said, we remark that, as expected from Fig. \ref{fig:everyxp}, the late-time features of the tracer dynamics are roughly similar, regardless the presence or absence of condensation. As done in the previous sections, we provide a comparison between the turbulent diffusion coefficient obtained from the tracers analysis and its theoretical counterpart, as predicted by the $K-\epsilon$ model. In this case, however, the introduction of the X-point magnetic geometry alters the structure and isotropy of the dynamical system (as it can be observed from \eqref{eqn}, \eqref{eqvort} and \eqref{eqohm}), so that the latter is no longer isomorphic to a 2D Euler equation. It is therefore straightforward to understand that the estimate provided by the $K-\epsilon$ model is, in this case, less accurate. For what concerns the condensation case, we calculate a typical size of eddies $L \simeq 11.2$, leading to an estimate for the turbulent diffusion coefficient $\mathcal{D}_T\simeq 0.037$. In the absence of condensation, instead, we deduce from the tracers data a typical size of turbulence $L\simeq 4.1$, which in turn implies $\mathcal{D}_T\simeq 0.01$. In both cases the theoretical paradigm offered by the $K-\epsilon$ model provides acceptable estimates for the turbulent diffusion coefficients obtained from the tracers motion, namely the values calculated from data amounts to roughly half the corresponding estimates.
\begin{figure}[hbtp!]
\centering
\includegraphics[width=6cm]{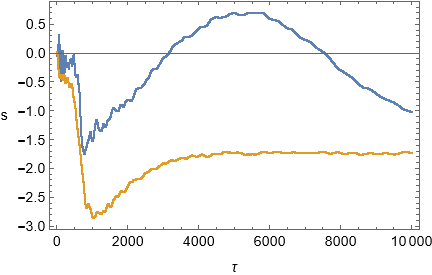}\!\!
\includegraphics[width=6cm]{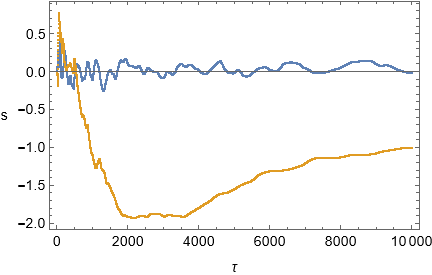}
\caption{Case B. Skewness: condensation (left) and non-condensation (right). The blue curve describes the skewness along the $u$ direction, orange color for the $v$ direction.}
\label{fig:skewpx}
\end{figure}
\begin{figure}[hbtp!]
\centering
\includegraphics[width=7cm]{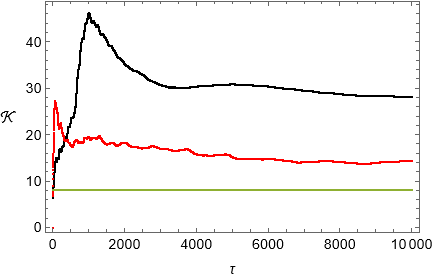}
\caption{Case B. Kurtosis: the condensation case is described by the black curve, non-condensation in red. The green line indicates the reference value for a Gaussian distribution.}
\label{fig:kurtpx}
\end{figure}

Let us now consider the normality tests on the distribution function. We can observe that, also in this aspect, the X-point magnetic field produced significant differences with respect to previous cases. First we comment on the symmetry test conducted via the computation of skewness, whose plots are given in Fig. \ref{fig:skewpx}. It is immediate to notice that the assumption of an X-point configuration introduces a significant amount of anisotropy. Indeed, as it can be observed in both the left and right panel, the blue and the orange curves, describing skewness in the $u$ and $v$ directions, respectively, have two completely different behaviors. For instance, if we focus on the right panel we see that the skewness on the $u$ direction is practically negligible. The same does not hold for the correspondent quantity along the $v$ direction, the latter presenting a significant deviation from normality. As previously stated, the precise outcome of this test has a considerable dependence from the particular choice of initial data for the electric potential simulation, given that the latter generates a peculiar pattern of vortices affecting the fluid tracers dynamics and, therefore, a specific realization of anisotropies. 
\begin{figure}[hbtp!]
\centering
\includegraphics[width=7cm]{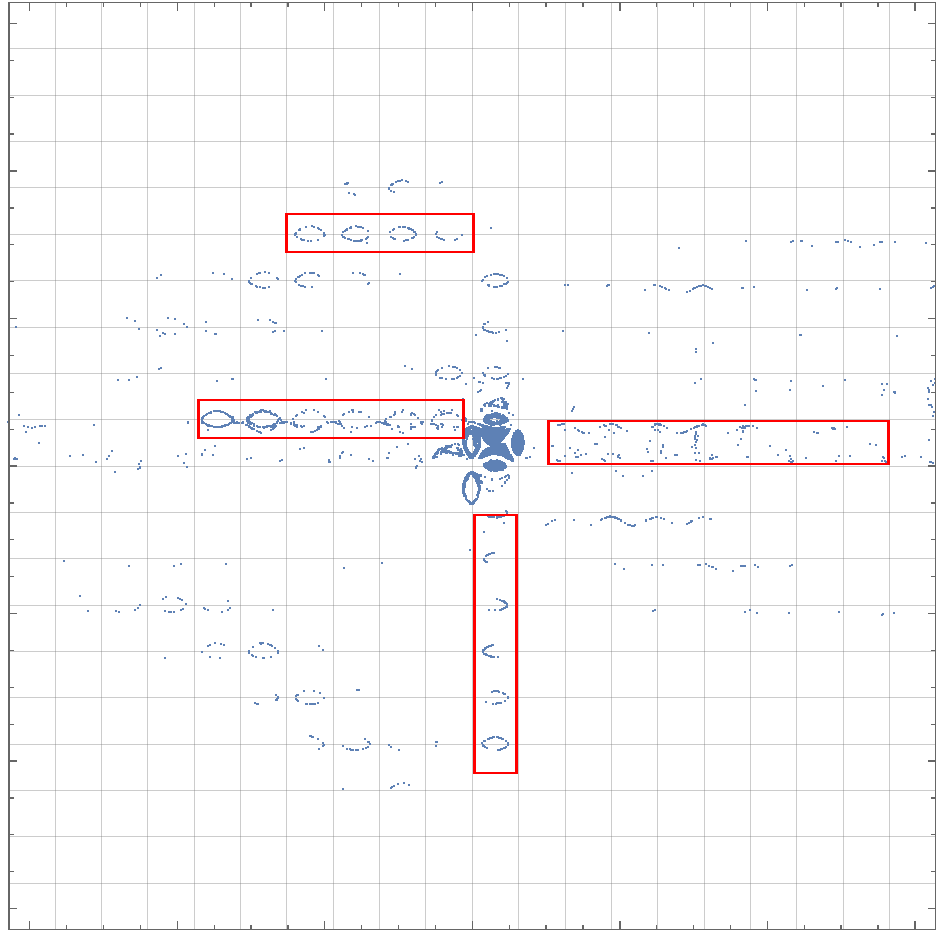}
\caption{Case B. Snapshot of the tracers population at the end of the simulation time in the poloidal plane $(u,v)$. The trajectories along which the outliers are transported much further than the average, here dubbed ``streams'', are highlighted by red boxes.}
\label{fig:snapfinale}
\end{figure}

Last, we display the curves describing the estimator of kurtosis, in Fig. \ref{fig:kurtpx}. We outline two main differences with the previous sections. First, the deviation from the Gaussian value of $8$ is much greater in this case: when we considered a toroidal magnetic field we calculated a deviation from normality roughly equal to $1$, whereas here we notice that the departure has a magnitude up to circa $40$. Moreover, contrarily to previously analyzed scenarios, the difference between the kurtosis calculated from the tracer data and the standard Gaussian value of $8$ is, in this case, positive. This implies that the distribution function of the tracer positions is strongly leptokurtic, with tails much more heavily populated than a Gaussian with equal variance. Hence, we observe that the introduction of the X-point configuration resulted in both a greater confinement, signaled by smaller values of the MSD, and in the generation of a large population of outliers, whose positions are much further from the initial ones with respect to the average. These two features seem, at first glance, somewhat counterintuitive and even contradictory, but an explanation for both phenomena can be provided. As previously stated, the decrease of the MSD is due to the change in topology induced by the X-point, leading to limited and disconnected regions and, on average, on a mitigated transport. Nevertheless, it is possible to observe that there are few specific trajectories along which the tracers migrate far from the initial positions. These particular curves on the poloidal plane are able to connect different regions thanks to the periodic boundary condition characterizing our numerical integration. The landscape can be described as a small number of narrow streams advectively transporting few portions of fluid with great efficiency. In Fig. \ref{fig:snapfinale} we report a snapshot of the tracer population, taken at the end of the simulation time. From this picture it can be observed that a small portion of markers is transported much more than the average along the aforementioned privileged trajectories. A comparison between the black and red curves reported in Fig. \ref{fig:kurtpx} shows that a greater deviation from Gaussianity is more marked in the presence of condensation, accordingly to the different morphology of the energy spectrum at low $k$ values with respect to the non condensation case.

\section{Concluding remarks}

In the analysis above, we performed a detailed investigation about the 
influence that the two-dimensional 
electrostatic turbulence exerts on the test charged particle motion, 
according to the $\mathbf{E}\times \mathbf{B}$ drift velocity, as in \cite{2023JPlPh..89a9008S,2007JNuM..363..550C}. 
The tracers considered in this work can clearly represent both 
plasma constituents as well as 
other ion species living in the 
edge region of a Tokamak device. We discussed the slab magnetic configuration in the inviscid and in the viscous case and also in correspondence to two 
different spectral setups, 
i.e. with and without condensation phenomenon. 
Then, we compared these results 
with the important case of a magnetic configuration in which the poloidal field well mimics a real X-point in a 
Tokamak equilibrium. 

The slab shaped case outlined 
that an inverse energy cascade (the condensation process typical of fluid dynamics) induces a more efficient transport characterized by both diffusive and anomalous phases, in contrast to a direct enstrophy cascade, i.e. when condensation is absent, the latter being responsible for a quasi-pure diffusive transport of smaller magnitude. 
This result can be interpreted, from a physical point of view, in the capability of a condensation process to 
generate large scale eddies, 
able to inhibit the stochastic 
field-particle scattering, while 
inducing a kind of particle trapping. Nonetheless, the emergence of larger spatial scales in the condensation case implies also an enlargement of the mean free path for particles interacting with vortices, resulting in an overall greater spreading of the tracer population.  

The results of the analysis with 
an X-point magnetic configuration show a reduction of the diffusivity by a factor of roughly $1/3$, in comparison to the 
pure slab case. This is due to the suppression of the turbulence intensity by means of the magnetic shear that the X-point configuration 
unavoidably induces \cite{Scott_2007}. 
However, a very intriguing feature emerges when the 
tracer statistical distribution function is analyzed. In fact, while the 
variance of such a distribution is, as expected from a lower diffusivity (smaller than in the slab 
case), the corresponding kurtosis is now much greater than its 
expected Gaussian value 
(well-recovered in the slab case). 
This feature has the important 
consequence that, if the tracers 
explore the X-point region, 
a small population of them is 
able to perform an efficient transport, well-above the corresponding 
diffusivity. 

Summarizing, the merit of the present study is to fix the following 
two relevant claims:
i) in the slab case, the presence of the condensation phenomenon for the turbulence spectrum implies a larger value of the diffusion coefficient, where it can be recovered, with respect to the case of a direct enstrophy cascade. Nonetheless, there is 
an enhancement of the asymptotic value of the MSD for this former case, estimated in about $50 \%$, 
so that a marked transport process is recovered.
ii) in the comparison between the 
presence and absence of an X-point configuration, typical of a real Tokamak device \cite{snowflake15},
we notice a  
smoother transport process, i.e. a smaller asymptotic value of the MSD. 

The first result can be interpreted by noticing that the formation of large 
sized eddies, in the case of a condensation of the spectrum, causes a growth of the mean free path for tracers between interactions with vortices and this results, on average, in a greater amount of transport.  
The presence of the X-point and then the emergence of shear magnetic contributions is instead at the ground of the second claim. In fact, such shear 
terms reduce the turbulence of 
the fluctuating fields and favor 
a smaller asymptotic MSD. 
However, it has to be stressed 
that, looking at the statistical distribution of the tracers, we see that, although the variance of the distribution is consistently smaller than the one in the slab case, the kurtosis of such a distribution is correspondingly larger. This feature suggests that the X-point configuration reduces, on average, the transport process, but enhances the number of outliers that can 
efficiently escape from the 
initial position. 

It is also remarkable the agreement that we find between the diffusion coefficient obtained from the tracers analysis and the one we estimate from the 
$K - \epsilon$ model, both in the case of 
a slab configuration and in the presence of an X-point magnetic configuration. This fact suggests that our investigation is validated in its predictivity by the capability to reproduce the right turbulence feature of the two-dimensional model 
we adopted here. 
In this respect, we stress that, although 
we consider a reduced model for 
the electric field dynamics (according to the analysis in \cite{2023PhyD..45133774M}), its 
non-linear features are just those of a fully self-sustained non-linear drift response. 
By other words, our reduced model properly captures all the important properties of the emerging turbulence and the confirmation of its viability is just 
in the correct prediction of the 
diffusion coefficient, that such a turbulent regime would generate according to the $K-\epsilon$ model.

We see how the present analysis 
offers intriguing hints on what 
should be an optimization of 
the turbulence features to deal 
with a well-behaving (confined) 
plasma: we should, first of all,  reduce the impact of the 
condensation phenomenon, so that mainly small scale eddies are formed lowering the magnitude of transport. 
Then, it is rather convenient 
if the particles near enough to the 
X-point are prevented to reach the 
external region by the creation of a transport barrier. In fact, the population of outliers, which are 
able to have larger values of 
the MSD, should be somehow 
preventively confined. 

In the real picture of a Tokamak, this 
could be achieved either by the introduction of a non-fluctuating radial electric field \cite{doi:10.13182/FST12-A13504,OSORIOQUIROGA2023100023}, or by an efficient 
transfer of particles from the 
X-point to the limiter, inhibiting the condensation of the turbulence 
spectrum. In the limiter region, 
the magnetic shear is much smaller 
than in the X-point region and 
the diffusivity of a non-condensed spectrum could ensure a significant 
impact on the outgoing particles.


\end{document}